\documentclass{aa}
\usepackage{graphicx}
\usepackage{natbib}
\usepackage{txfonts}
\newcommand{\h}{$^{\rm h}$}
\newcommand{\m}{$^{\rm m}$}

\begin{document}
   \title{Size and properties of the narrow-line region in Seyfert-1 galaxies
from spatially-resolved optical spectroscopy
\thanks{Based on observations made with ESO Telescopes at
   the Cerro Paranal Observatory under programme ID 72.B-0144 and the La Silla Observatory under programme ID 073.B-0013}}

   \author{Nicola Bennert\inst{1,2}
          \and
           Bruno Jungwiert\inst{2,3,4} \and Stefanie Komossa\inst{5} \and
          Martin Haas\inst{1} \and Rolf Chini\inst{1}
          }

   \offprints{Nicola Bennert}
\institute{Astronomisches Institut Ruhr-Universit\"at Bochum,
              Universit\"atsstrasse 150, D-44780 Bochum, Germany;
\email{haas@astro.rub.de}, \email{chini@astro.rub.de}
\and
Institute of Geophysics and Planetary Physics, University of California, Riverside, CA 92521, USA;
\email{nicola.bennert@ucr.edu}
\and
Astronomical Institute, Academy of Sciences of the Czech Republic,
Bo{\v c}n\'\i\ II 1401, 141 31 Prague 4, Czech Republic;
\email{bruno@ig.cas.cz}
          \and
             CRAL-Observatoire de Lyon, 9 avenue Charles Andr{\'e}, F-69561 
Saint-Genis-Laval cedex, France
\and
   Max-Planck Institut f\"ur extraterrestrische Physik,
              Giessenbachstrasse 1, D-85748 Garching, Germany; \email{skomossa@xray.mpe.mpg.de}
}

   \date{Received; accepted}


   \abstract{Spatially resolved emission-line spectroscopy
is a powerful tool to determine the physical conditions in the
narrow-line region (NLR) of active galactic nuclei (AGNs).
We recently used optical long-slit spectroscopy
to study the NLRs of a sample
of six Seyfert-2 galaxies. We have shown that such an approach, 
in comparison to the commonly used [\ion{O}{iii}] narrow-band imaging alone,
allows us to probe the size of the NLR in terms of AGN photoionisation.
Moreover, several physical parameters of the NLR can be directly accessed.}
{We here apply the same methods 
to study the NLR of six Seyfert-1 galaxies and compare our results
to those of Seyfert-2 galaxies.}
{We employ diagnostically
valuable emission-line ratios to determine the physical
properties of the NLR, including the core values and radial
dependencies of density, ionisation parameter, and reddening.
Tracking the radial change of emission-line ratios in diagnostic
diagrams allows us to measure the transition between AGN-like and
\ion{H}{ii}-like line excitation, and thus we are able to measure the size
of the NLR.}
{In the diagnostic diagrams,
we find a transition between line ratios falling in the
AGN regime and those typical for \ion{H}{ii} regions in two
Seyfert-1 galaxies, thus determining the size of the NLR.
The central electron temperature and ionisation parameter
are in general higher in type-1 Seyferts than in type 2s. 
In almost all cases,
both electron density and ionisation parameter decrease with radius.
In general, the decrease is faster in Seyfert-1 galaxies than in type 2s.
In several objects, the gaseous velocity distribution is characteristic for
rotational motion in an (inclined) emission-line disk in the centre.
We give estimates of the black hole masses.
We discuss our findings in detail for each object.}
{}

\keywords{Galaxies: active --
          Galaxies: nuclei --
          Galaxies: Seyfert}
\titlerunning{The NLR in Seyfert-1 galaxies}
\authorrunning{N. Bennert et\,al.}

   \maketitle
%

\section{Introduction}
The cores of active galaxies, most likely powered
by accretion onto supermassive black holes (BH), are surrounded by two
emission-line regions; the broad-line region (BLR) in proximity
to the BH, and the
narrow-line region (NLR) at larger distances from the nucleus. The study of
both BLR and NLR
provides us with important information on the nature and origin
of these cloud systems, on their link with the host galaxy, and on
their cosmological evolution. Determination of their velocity field
and distance from the nucleus also enables us to estimate BH
masses. While the BLR is too close to the nucleus, the NLR of many
AGN is spatially resolved, and we can thus extract information on the
NLR properties by performing spatially resolved spectroscopy.
This method is a powerful approach
to measure the physical conditions in the NLR  and surrounding
regions [e.g.~\citet{wil89, rob94, rad98, sch99, fra00, bar01, sos01, tem03, cir05}].

While [\ion{O}{iii}]\,$\lambda$5007\AA~(hereafter [\ion{O}{iii}])
narrow-band imaging is commonly used to study the NLRs of active
galaxies, we have shown in \citet{ben06a} and \citet{ben06b} 
(hereafter paper I \& II) that 
this emission can be contaminated by contributions from
star formation and that
different sensitivities can lead to 
different size measurements of the NLR.
Using long-slit spectroscopy, 
we developed methods to 
probe the AGN-photoionisation of the NLR and thus, its ``real'' size.
From spatially resolved spectral diagnostics,
we find a transition between central line ratios falling into the
AGN regime and outer ones in the \ion{H}{ii}-region regime for two objects. 
Applying \texttt{CLOUDY} photoionisation models \citep{fer98}, we show that the observed distinction
between \ion{H}{ii}-like and AGN-like ratios represents a true
difference in ionisation source and cannot
be explained by variations of physical parameters such as
ionisation parameter, electron density or metallicity. We 
interpret it as a real border between the NLR,
i.e. the central AGN-photoionised region, and 
surrounding \ion{H}{ii} regions.
In addition, several physical parameters of the NLR such as reddening, ionisation parameter,
electron density, and velocity can be directly
accessed and analysed as a function of distance from the nucleus.
We find that both the electron density and the ionisation parameter decrease with
radius. 
The differences between the reddening distributions 
determined from the continuum slope and the Balmer
decrement argue in favour of dust intrinsic to the NLR clouds with 
varying column density along the
line of sight.
The NLR and stellar velocity fields are similar and indicate that
the NLR gas is distributed in a disk rather than a sphere.

Here, we apply the same methods
to a sample of six Seyfert-1 galaxies 
to probe the size of the NLR and
derive physical properties such as reddening, ionisation parameter,
electron density, and velocity in type-1 AGNs. We discuss their
variations with distance from the nucleus and
compare the results for Seyfert 1s and Seyfert 2s, allowing
to test the facets of the unified model of AGNs.
A detailed comparison
of our results with literature data is given for each object [Appendix; see also \citet{ben05}].

\section{Observations, Reduction, and Analysis}
\begin{figure*}
\begin{center}
\includegraphics[width=16cm]{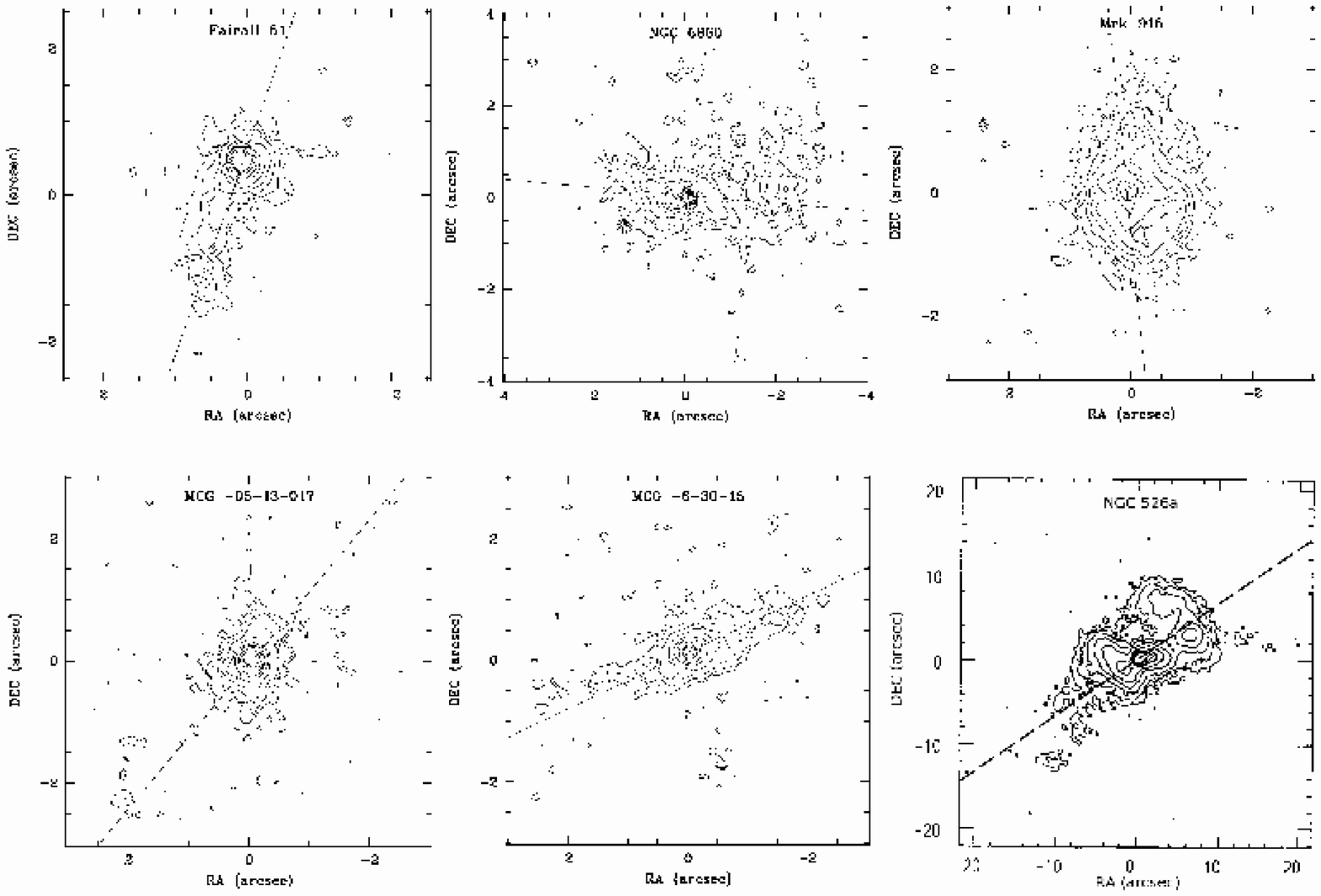}
\caption[]{\label{galaxies1} \small 
HST [\ion{O}{iii}] images of  Fairall\,51, NGC\,6860, Mrk\,915, MCG\,-05-13-017, and MCG\,-6-30-15 taken
from \citet{sch03a} (WF chip: $\sim$0\farcs1\,pix$^{-1}$).
Contours start at the 3$\sigma$ level above the background 
(\citet{sch03a}, their Table 2) and increase in
powers of 2 times 3$\sigma$ (3$\sigma$ $\times$ $2^n$). For NGG\,526a, a
groundbased image taken from \citet{mul96a} is shown.
The position of the long slit is shown as dashed line.
North is up, east to the left.
}
\end{center}
\end{figure*}

The spectra were obtained with FORS1@VLT and EMMI@NTT.
Relevant information on the sample and observations is summarised in
Tables~\ref{objsy1} and~\ref{obssy1}.
The [\ion{O}{iii}] images with the slit position overlaid are shown in
Fig.~\ref{galaxies1}.

As the observations, reduction, and analysis were already described in detail in paper II,
we here discuss the special treatment of the Seyfert-1 spectra only.

\begin{table*}
\begin{minipage}{180mm}
 \caption[]{\label{objsy1} Properties of the sample\footnote{Unless stated otherwise, the properties were taken from
the NASA/IPAC Extragalactic Database (NED).}}
\begin{center}
\begin{tabular}{lcccccc}
\\[-2.3ex]
\hline
\hline\\[-2.3ex]
& Fairall\,51 & NGC\,6860 & Mrk\,915 & NGC\,526a & MCG\,-05-13-017 & MCG\,-6-30-15\\[0.25ex]
\hline\\[-2.3ex]
altern. name & ESO\,140-G043 & ESO\,143-G009 & MCG\,-02-57-023 &
MCG\,-06-04-019 & ESO\,362-G018 & ESO\,383-G035\\
$\alpha$ (J2000) & 18\h44\m54\fs0  & 20\h08\m46\fs9
&  22\h36\m46\fs5 & 01\h23\m54\fs4 
&  05\h19\m35\fs8 &  13\h35\m53\fs8\\
$\delta$ (J2000) & -62\degr21\arcmin53\arcsec &  -61\degr06\arcmin01\arcsec
&   -12\degr32\arcmin43\arcsec & 
-35\degr03\arcmin56\arcsec &  -32\degr39\arcmin28\arcsec & -34\degr17\arcmin44\arcsec\\
i. (\degr)\footnote{Host galaxy inclination [\citet{vau91}; RC3]} & 64 & 61 & 80 & 55 & 54 &  57\\
p.a. (\degr)\footnote{Position angle of host galaxy major axis (RC3)}& 162 & 34 & 166 & 112 & 160 &  116\\
$v_{\rm hel}$ (km\,s$^{-1}$) & 4255$\pm$10 & 4462$\pm$24 & 7228$\pm$2
& 5725$\pm$39 & 3790$\pm$30 & 2323$\pm$15\\
$v_{\rm 3k}$ (km\,s$^{-1}$)\footnote{Velocity relative to the 3K background using the NED velocity calculator}& 4228 & 4377 & 6863 & 5446 & 3620 & 2595\\
dist. (Mpc)\footnote{Distance $D$ in Mpc, using $v_{\rm 3K}$ and $H_0$ = 71\,km\,s$^{-1}$\,Mpc$^{-1}$}& 60 & 62 & 98 & 78 & 52 &  37\\
lin. scale (pc/\arcsec)\footnote{Linear scale $d$ using distance $D$ and $d$ = 4.848 $\cdot$ 10$^{-6}$
$\cdot$ $D$}& 283 & 293 & 454 & 362 & 243&   175\\
morphology & (R'\_2)SB(rs)b & (R')SB(r)ab & Sb & S0pec? & S0/a&  E-S0\\
AGN type (NED) & Sy1 & Sy1 & Sy1 & Sy1.5 & Sy1.5 &  Sy1.2\\
AGN type (our spectra) & Sy1 & Sy1.5 & Sy1.5 & Sy1.9 & Sy1.5 & Sy1.2\\
$E_{\rm (B-V),G}$ (mag)\footnote{Foreground Milky Way reddening used for reddening correction \citep{sch98}} & 0.108 & 0.041 & 0.063 & 0.028 & 0.017 &  0.062\\
$M_B$ (mag) & 14.7 & 13.68 & 14.82 & 14.5 & 12.5&  13.7\\[0.1ex]
\hline\\[-2.3ex]
\end{tabular}
\end{center}
\end{minipage}
\end{table*}

\begin{table*}
\begin{minipage}{180mm}
 \caption[]{\label{obssy1} Observations of the sample}
\begin{center}
\begin{tabular}{lcccccc}
\\[-2.3ex]
\hline
\hline\\[-2.3ex]
& Fairall\,51 & NGC\,6860 & Mrk\,915 & NGC\,526a & MCG\,-05-13-017 & MCG\,-6-30-15\\[0.25ex]
\hline\\[-2.3ex]
telescope & NTT & NTT & NTT & NTT & NTT & VLT\\
date (beg.) & 14/15-Sep-04 & 15/16-Sep-04  & 14/15-Sep-04 & 16-Sep-04 & 17-Sep-04&  25-Feb-04 \\
exp. time  blue (s)\footnote{Total integration time. At the VLT, 
the blue and red spectral range were covered in one exposure.} & 3000 & 6000 & -\footnote{The spectra taken in the blue wavelength range were corrupted
due to instrumental problems.}
 & 3000 & 3600 & 1800\\
exp. time  red (s)$^a$  & 3600 & 3600 & 3000 & 2400 & 3000 & 1800\\
seeing & $<$ 1\arcsec & $<$ 1\arcsec & $<$ 1\arcsec & $<$ 1\arcsec & $<$ 1\arcsec & $\sim$1\farcs5\\
slit width & 1\arcsec & 1\arcsec & 1\arcsec & 1\arcsec & 1\arcsec& $\sim$1\arcsec\\
FWHM$_{\rm instr}$ (km\,s$^{-1}$) & 250 & 250 & 250 & 250 & 250 & 590\\
p.a.~(\degr)\footnote{Position angle of the slit} & 160 & 85 &  5 & 123 & 140 &   115\\
hel. corr. (km\,s$^{-1}$)\footnote{This heliocentric correction was added to the measured radial
velocities.}  & -2 & 0 & -4 & +17 & +29 &  +12\\
average (pixel)\footnote{Number of pixel rows which were averaged} & 3 & 3 & 3 & 3 & 3 &  7\\
scale\footnote{Formal spatial resolution of final extracted spectra}&  1\farcs1 $\times$ 1\arcsec& 1\farcs1 $\times$ 1\arcsec &
1\farcs1 $\times$ 1\arcsec & 1\farcs1 $\times$ 1\arcsec & 1\farcs1
$\times$ 1\arcsec &  1\farcs4~$\times$ 1\arcsec\\[0.1ex]
\hline\\[-2.3ex]
\end{tabular}
\end{center}
\end{minipage}
\end{table*}

\subsection{Subtracting the stellar population}
\label{stellarpop}
As discussed in paper I \& II, removing the contribution of the
stellar population is one of the first and most critical steps in the
analysis of AGN emission-line spectra, at least in Seyfert-2 galaxies. 

For Seyfert-1 galaxies, the procedure described in paper I \& II 
may not be simply applicable:
The AGN featureless continuum can be very strong, especially
in the central parts where the broad emission lines are seen.
Thus, a stellar template cannot simply be scaled to the continuum value
in these regions as the contribution of the underlying stellar population would
be overestimated. 

However, for type-1 AGNs, the AGN continuum 
and the broad and narrow emission lines often completely dominate the spectrum.
We did not find signs of strong underlying Balmer absorption lines. In some cases,
faint absorption is visible in \ion{Ca}{ii} H\&K and in \ion{Na}{i} D. 
Some of the \ion{Na}{i} D absorption may be of interstellar origin.
We consider the underlying stellar absorption as negligible compared
to the central emission line fluxes.
Moreover, it was difficult to derive a suited high S/N
template.

\begin{figure*}  
\includegraphics[width=18cm]{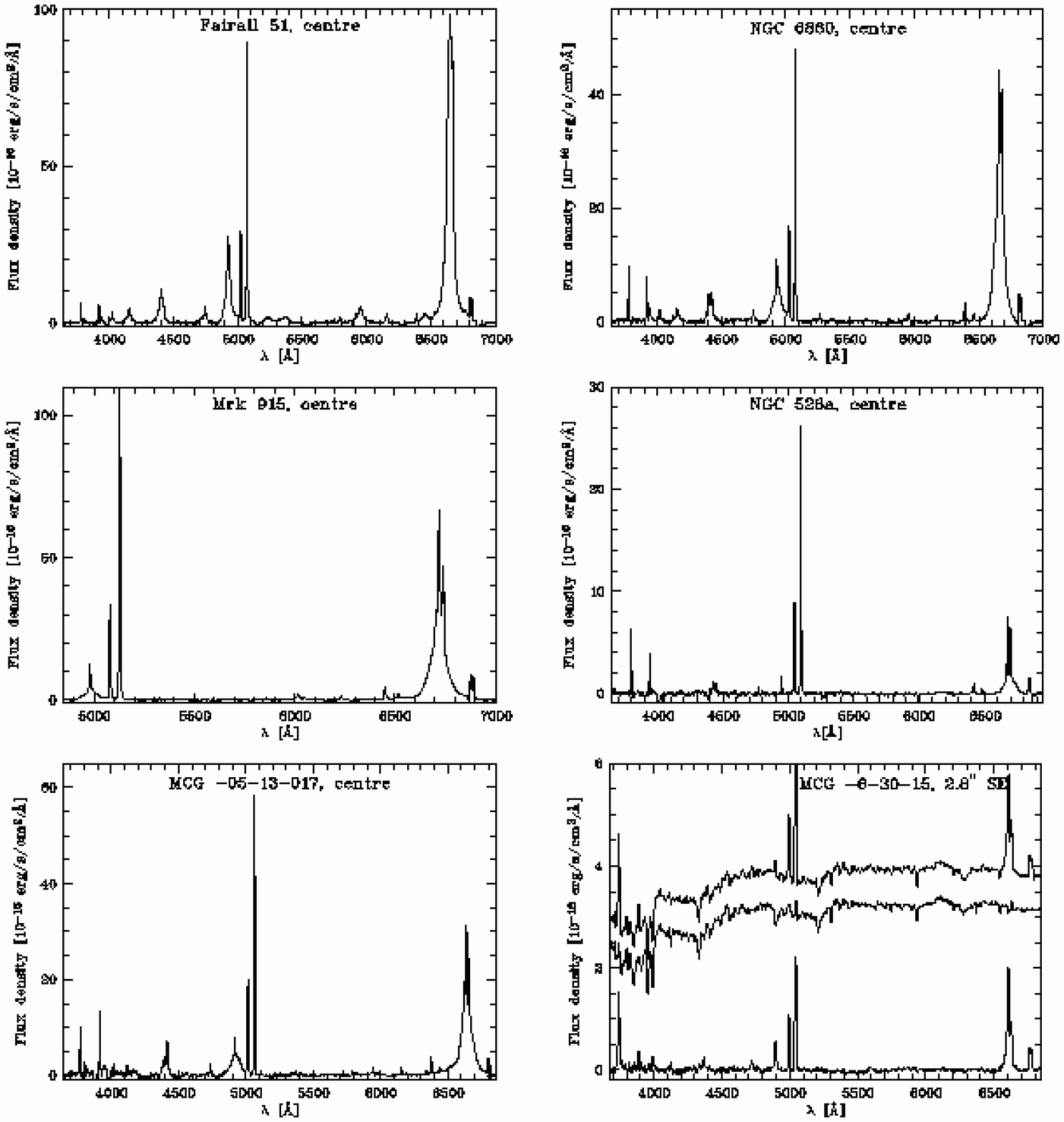}
\caption[Central spectra]
{\label{figtemplate} \small
Spectra of the six Seyfert-1 galaxies in our sample.
For MCG\,-6-15-30, we show the
template subtraction: The template was obtained
at 12\arcsec~north-west of the nucleus, averaged over 2\arcsec
and median-filtered over three pixel to increase the S/N.
The observed (upper; at 2\farcs8 south-east from the centre), 
the template (middle) and
the template-subtracted spectrum (lower spectrum) are shown.
In this plot, both upper spectra are shifted
vertically by an arbitrary amount.
Strong emission lines are truncated in the difference spectrum.
The template matches the stellar absorption lines seen in the NLR spectrum
fairly well.}
\end{figure*}

Only for MCG\,-6-30-15, a correction of the stellar population
was both necessary and possible.
We were able to gain a suited template free of
contaminating emission lines (Fig.~\ref{figtemplate}).
Thus, a correction of underlying stellar absorption lines in the Seyfert-1 galaxies
were applied only to MCG\,-6-30-15. We scaled the template to the continuum
as we do not know the contribution of a featureless continuum.
The reddening measure using the continuum
slope variation relative to the stellar template (see paper I \& II) was, among the Seyfert-1
galaxies, only determined for MCG\,-6-30-15.

\subsection{\ion{Fe}{ii} contamination}
When studying optical spectra of type-1 AGNs,
another issue that needs to be taken into account is the contribution
of broad \ion{Fe}{ii} emission.
To probe the contribution of \ion{Fe}{ii} to the observed type-1 spectra,
we used the \ion{Fe}{ii} template of \citet{ver04}.
It was rebinned to the same resolution and shifted to the object's
redshift. We used several scaling factors and subtracted the template.
The residual continuum was searched for signs of remaining
\ion{Fe}{ii} emission. However, in all our type-1 objects,
the contribution of \ion{Fe}{ii} seems to be negligible
and for most scalings, we artificially induced ``\ion{Fe}{ii} 
absorption lines'', indicating that the scaling was too high.
Thus, no \ion{Fe}{ii} template was finally subtracted as we believe that the
\ion{Fe}{ii} contribution is negligible in our Seyfert-1 galaxies.

\subsection{Emission-line fluxes and reddening}
To determine the fluxes of the pure emission-line spectra,
the same general procedure as for the Seyfert-2 galaxies
described in paper I \& II was applied.

However, for the Seyfert-1 galaxies discussed here,
the fitting procedure is more difficult,
due to the additional broad lines of the BLR underlying all permitted emission
of the NLR. Broad H$\alpha$ lines are observed in the central spectra of 
all our type-1 objects (according to
their classification as Sy1, Sy1.2 or Sy1.5). 
Broad H$\beta$ emission is seen in all type 1s with
the exception of NGC\,526a, classifying it as Sy1.9 (see also Section~\ref{ngc526a}).

A common approach to disentangle the narrow and broad permitted lines is
to use the profile of the forbidden narrow lines such as [\ion{O}{iii}] as
template for the permitted narrow line, scaled to the appropriate height.
A second Gaussian (a broad one) is additionally used to fit the permitted
broad line profile.
During the fitting procedure, we found that the use of two Gaussians, a
broad and a narrow one, was in most cases not suited to fit the broad wings.
For all lines with underlying broad emission,
we added a third Gaussian: We fitted a narrow Gaussian, one
with an intermediate width,
and a broad one for an optimal total fit to the observed emission-line
profiles. 

Three Gaussians have already been used by
other authors to fit the H$\alpha$ and H$\beta$ lines in Seyfert-1 galaxies [e.g.~\citet{rey97,sul02}]. 
Emission-line profiles represent line-of-sight integrations of several
kinematic components and even for ``narrow'' lines, considerable profile
structure is measured at sufficient resolution
[e.g.~\citet{vrt85,whi85,schu03}].
Gaussian fits or Lorentz
fits are commonly used when single component fits fail. \citet{whi85} already
describe the non-Gaussian nature of observed [\ion{O}{iii}] line profiles
``which revealed a stronger base relative to the core than
Gaussians''. A  Lorentz profile which has broader wings compared to a Gaussian
seems to be better suited as has been shown by \citet{ver01} for the 
broad emission lines in
narrow-line Seyfert-1 galaxies and by \citet{schu03} for narrow emission lines in
Seyfert-2 galaxies. \citet{ben04} suggest the use of $d$-Lorentzians which
allow to fit both permitted and forbidden lines by adjusting an additional
parameter $d$.
\citet{sul02} studied the broad H$\beta$ line in several AGN types and found that 
objects with full-width at half maximum (FHWM) $<$ $\sim$4000\,km\,s$^{-1}$ are well fitted by a Lorentz function, 
while AGNs with FWHM $>$ $\sim$4000\,km\,s$^{-1}$ 
are better fitted if two broad-line components are used: 
A ``classical'' broad-line component and a very broad/redshifted component.
Our results are in agreement with this trend: 
All objects with broad emission lines in both H$\alpha$ and
H$\beta$ have FWHM $>$ $\sim$4000\,km\,s$^{-1}$ and had to be fitted by two broad-line components.

To conclude, we used single Gaussians
to fit the narrow lines and three Gaussians to fit narrow lines with
underlying broad emission which yields a very good result, taking into account the
low resolution of our spectra.

In all but one case (MCG\,-6-30-15), 
the permitted profiles show a clear
separation between broad underlying emission and a narrow ``peak''.
Thus, for most Seyfert-1 galaxies, 
we were able to distinguish between the broad and narrow emission
using three Gaussians with one resembling the shape of forbidden narrow lines.
For MCG\,-6-30-15, the only type-1 observed with the lower
resolution of VLT/FORS1, the profile fitting to the permitted Balmer lines could not
successfully disentangle the broad and narrow line. Thus,
we applied a Gaussian fit to forbidden lines only (except
for the [\ion{N}{ii}]\,$\lambda\lambda$6548,6583\,\AA~lines 
which are blended by H$\alpha$).
In the central spectra, the broad emission of H$\beta$ and H$\alpha$ 
even affect the adjacent [\ion{O}{iii}] and 
[\ion{S}{ii}]\,$\lambda\lambda$6716,6731\,\AA~lines.
In those cases, we subtracted the broad underlying wing by extrapolation.
As a consequence, the only emission-line ratio we were able to derive directly
is that of the two forbidden sulphur lines to measure
the electron density. The narrow H$\alpha$ and
H$\beta$ emission-line fluxes are needed to plot diagnostic line-ratio
diagrams, thus we cannot present these results for MCG\,-6-30-15.
Moreover, the ionisation parameter strongly depends on the reddening value. As we cannot
estimate it from the narrow H$\alpha$/H$\beta$ ratio, we used as a first guess 
the reddening
slope determined by matching the stellar template to the NLR spectra.

\section{Results and Discussion}

\subsection{Nuclear spectra}
The central spectra of the galaxies in our sample are shown in Fig.~\ref{figtemplate}.
Table~\ref{lineratio1} lists the observed and
reddening-corrected line-intensity ratios relative to H$\beta$ from the
nuclear spectrum (uncorrected for slit losses). 
For pairs of lines 
([\ion{O}{iii}], [\ion{O}{i}], and [\ion{N}{ii}]) with a fixed
line ratio ($\sim$3:1), only the brighter line is used.
(Note that all ratios correspond to narrow lines.)
Emission-line ratios of the strongest (narrow) lines as a function of distance
from the centre can be found online for each individual galaxy (excluding
MCG\,-6-30-15 as we
were not able to disentangle the broad and narrow Balmer lines).

\begin{table*}
\begin{minipage}{180mm}
 \caption[]{\label{lineratio1} Observed and reddening-corrected 
narrow emission line intensity ratios relative to H$\beta$\footnote{All narrow emission line ratios were derived from the nuclear
spectra. After reddening correction, other Balmer line-ratios such as H$\gamma$/H$\beta$ and
H$\delta$/H$\beta$ are consistent with the recombination values within the
errors.  No ratios are given for MCG\,-6-30-15 as we
were not able to disentangle the broad and narrow Balmer lines in the central spectra.
The uncertainties are in the range of $\sim$1-15\%.}}
\begin{center}
\begin{tabular}{lcccccccccccc}
\\[-2.3ex]
\hline
\hline\\[-2.3ex]
\multicolumn{1}{c}{Line}
& \multicolumn{2}{c}{\rm Fairall\,51} & \multicolumn{2}{c}{\rm NGC\,6860}
& \multicolumn{2}{c}{\rm Mrk\,915} 
& \multicolumn{2}{c}{\rm NGC\,526a} & \multicolumn{2}{c}{\rm MCG\,-05-13-017}\\
& \multicolumn{1}{c}{$F_{\rm obs}$} & \multicolumn{1}{c}{$F_{\rm dered}$}
& \multicolumn{1}{c}{$F_{\rm obs}$} & \multicolumn{1}{c}{$F_{\rm dered}$}
& \multicolumn{1}{c}{$F_{\rm obs}$} & \multicolumn{1}{c}{$F_{\rm dered}$} 
& \multicolumn{1}{c}{$F_{\rm obs}$} & \multicolumn{1}{c}{$F_{\rm dered}$}
& \multicolumn{1}{c}{$F_{\rm obs}$} & \multicolumn{1}{c}{$F_{\rm dered}$}\\[0.25ex]
\hline\\[-2.3ex]
$[\ion{O}{ii}]\,\lambda3727$\,\AA & 0.94 & 1.21 & 1.60 & 2.04 & --\footnote{Not covered by wavelength range} & --$^b$ & 3.20 & 4.20
& 1.84 & 2.62 \\*[0.01cm]
$[\ion{Ne}{iii}]\,\lambda3869$\,\AA & 0.86 & 1.32 & 1.18 & 1.45 & --$^b$ & --$^b$
& 1.80 & 2.27 & 2.54 & 3.43 \\*[0.01cm]
$[\ion{Ne}{iii}]\,\lambda$3967\,\AA & 0.08 & 0.11 & 0.24 & 0.29 & --$^b$ & --$^b$
& --\footnote{Underlying absorption lines} & --$^c$ & 0.28 & 0.37 \\*[0.01cm]
$[\ion{O}{iii}]\,\lambda$4363\,\AA & 0.40 & 0.49 & 0.53 & 0.59 & --$^b$ & --$^b$ &
0.65 & 0.73 & 1.26 & 1.46\\*[0.01cm]
$\ion{He}{ii}\,\lambda$4686\,\AA & 0.41 & 0.44 & 0.29 & 0.30 & --$^b$ & --$^b$ &
0.33 & 0.34 & 0.48 & 0.50\\*[0.01cm]
$[\ion{O}{iii}]\,\lambda$5007\,\AA & \hspace*{-0.2cm}15.58 & \hspace*{-0.2cm}14.47 & 8.84 & 8.53 & \hspace*{-0.2cm}12.72 & \hspace*{-0.2cm}11.82 &
\hspace*{-0.2cm}16.35 & \hspace*{-0.2cm}15.71 & \hspace*{-0.2cm}17.15 & \hspace*{-0.2cm}16.29\\*[0.01cm]
$[\ion{Fe}{vii}]\,\lambda$5721\,\AA & 0.49 & 0.34 & 0.14 & 0.12 & 0.15 & 0.11 &
0.11 & 0.09 & 0.37 & 0.29\\*[0.01cm]
$[\ion{Fe}{vii}]\,\lambda$6087\,\AA & 0.83 & 0.49 & 0.20 & 0.16 & 0.27 & 0.16 & 0.09
& 0.07 & 0.47 & 0.33\\*[0.01cm]
$[\ion{O}{i}]\,\lambda$6300\,\AA & 0.92 & 0.51 & 0.71 & 0.53 & 0.81 & 0.45 & 0.74 &
0.54 & 1.19 & 0.79\\*[0.01cm]
$[\ion{Fe}{x}]\,\lambda$6375\,\AA & 0.81 & 0.43 & 0.08 & 0.06 & 0.17 & 0.09 & 0.06
& 0.04 & 0.38 & 0.25\\*[0.01cm]
H$\alpha$ &  5.60 & 2.87 & 3.95 & 2.87 & 5.60 & 2.87 & 4.11 & 2.87 & 4.57 & 2.87\\*[0.01cm]
$[\ion{N}{ii}]\,\lambda$6583\,\AA & 5.57 & 2.84 & 3.38 & 2.44 & 3.70 & 1.88 & 3.14 &
2.18 & 3.02 & 1.89\\*[0.01cm]
$[\ion{S}{ii}]\,\lambda$6716\,\AA & 1.47 & 0.73 & 1.28 & 0.92 & 1.29 & 0.64 & 1.18
& 0.81 & 1.01 & 0.62\\*[0.01cm]
$[\ion{S}{ii}]\,\lambda$6731\,\AA & 1.68 & 0.84 & 1.15 & 0.82 & 1.31 & 0.65 & 1.16
& 0.80 & 1.19 & 0.73\\*[0.01cm]
\hline\\[-2.3ex]
\end{tabular}
\end{center}
\end{minipage}
\end{table*}

In Table~\ref{result}, we give the reddening-corrected H$\beta$ luminosity
and summarise the results from dereddened line ratios such as the electron
temperature $T_{\rm e, obs}$\footnote{Derived from the [\ion{O}{iii}]($\lambda$4959\,\AA+$\lambda$5007\,\AA)/$\lambda$4363\,\AA~emission-line ratio}, 
the reddening value $E_{B - V}$, 
the electron density $n_{\rm e, obs}$, and the ionisation
parameter $U_{\rm obs}$ for the nuclear spectra of all objects.
The parameters represent an average over the central several hundred
parsecs.

The temperature was, in most objects, only
determined for the nuclear spectrum due to the faintness of the involved 
[\ion{O}{iii}]\,$\lambda$4363\,\AA~emission line in the outer spectra.
In two objects, we were able to derive the electron temperature in the inner
few arcseconds (NGC\,526a,
MCG\,-05-13-017) where it stays roughly
constant within the errors or scatters without showing a clear dependency on radius. 
The central temperature was used to apply a correction to the electron density.
In those cases in which no temperature was measured we used $T = 10000$\,K or
an average temperature derived from the other galaxies instead. 

\begin{table*}
\begin{minipage}{180mm}
\caption[]{\label{result} Reddening-corrected narrow H$\beta$ flux and
luminosity and 
results from dereddened narrow emission line ratios of the nuclear spectra.}
\begin{center}
\begin{tabular}{lcccccc}
\\[-2.3ex]
\hline
\hline\\[-2.3ex]
 & \rm{Fairall\,51} & \rm{NGC\,6860} & \rm{Mrk\,915} &  \rm{NGC\,526a} & \rm{MCG\,-05-13-017}& \rm{MCG\,-6-30-15} \\[0.25ex]
\hline\\[-2.3ex]
$F_{\rm H\beta}$ (10$^{-14}$\,erg\,s$^{-1}$\,cm$^{-2}$) & 21$\pm$2 & 8$\pm$0.5 & 36$\pm$3 & 3$\pm$0.1 & 8$\pm$0.6 & --\footnote{No deconvolution of underlying broad Balmer line possible}\\
$L_{\rm H\beta}$ (10$^{39}$\,erg\,s$^{-1}$) & 93$\pm$9                   & 37$\pm$2                   & 416$\pm$30                 & 18$\pm$1                    & 25$\pm$2& --$^a$                      \\
$T_{\rm e, obs}$ (K)                                                             & \hspace{-2mm}22200$\pm$400 & \hspace{-2mm}36325$\pm$250 & \hspace{+4mm}--\footnote{Not covered by wavelength range}         & 23330$\pm$1700              & 52500$\pm$3000& --$^a$ \\
$E_{B - V}$ (mag)\footnote{Note that this central value is not necessarily representative for the reddening within the NLR; for more details on reddening see Table~\ref{reddening}}                            & \hspace{+2mm}0.59$\pm$0.03 & \hspace{+2mm}0.28$\pm$0.02 & \hspace{+2mm}0.59$\pm$0.02 & \hspace{+2mm}0.32$\pm$0.03  & \hspace{+2.5mm}0.41$\pm$0.03& \hspace{+2mm}0.3$\pm$0.02\footnote{Determined from reddening of continuum slope relative to template} \\
$n_{\rm e, obs}$ (cm$^{-3}$)                     & \hspace{-2mm}1430$\pm$40   & \hspace{-2mm}1015$\pm$50   & 570 (1045)$\pm$35\footnote{[\ion{S}{ii}]\,$\lambda$6731\,\AA~is slightly truncated by telluric absorption bands.}$^,$\footnote{Using $T_e$ = 10000\,K and, in brackets, $<$$T_{e}$$>_{\rm 4 Sy1s}$ $\sim$
33590, respectively}
  & \hspace{+2mm}835$\pm$70$^e$ & \hspace{-2mm}2460$\pm$55& 300 (550)$\pm$40$^f$        \\
$U_{\rm log (n_e) = 3, obs}$  (10$^{-3}$)        & 9.25$\pm$0.9               & \hspace{+2mm}2.73$\pm$0.04 & \hspace{+4mm}--$^b$         & \hspace{+2mm}2.89$\pm$0.05  & 4.28$\pm$0.1& \hspace{-2mm}2.95$\pm$0.04    \\[0.1ex]
\hline\\[-2.3ex]
\end{tabular}
\end{center}
\end{minipage}
\end{table*}

\subsubsection{Comparison of Sy1 and Sy2 properties}
\label{longdiff12}
Comparing the results for the central spectra of 
type-1 and type-2 Seyferts (paper II) shows that
the line ratios are similar in all objects. There are no
significant differences between type-1 and type-2 galaxies with the
exception of the emission lines of oxygen and iron which are on average higher
in the Seyfert-1 galaxies: 
[\ion{O}{iii}]\,$\lambda$4363\,\AA/H$\beta$ $\sim$ 0.82$\pm$0.2 (4 Sy1s) versus
0.19$\pm$0.02 (4 Sy2s);
[\ion{O}{iii}]\,$\lambda$5007\,\AA/H$\beta$ $\sim$ 13.4$\pm$1.4 (5 Sy1s)
versus 10.3$\pm$0.6 (6 Sy2s); 
[\ion{Fe}{vii}]\,$\lambda$5721\,\AA/H$\beta$ $\sim$ 0.19$\pm$0.05
(5 Sy1s) versus 0.14$\pm$0.1 (2 Sy2s);
[\ion{Fe}{vii}]\,$\lambda$6087\,\AA/H$\beta$ $\sim$ 0.24$\pm$0.07 
(5 Sy1s) versus 0.14$\pm$0.06 (4 Sy2s);
[\ion{Fe}{x}]\,$\lambda$6375\,\AA/H$\beta$ $\sim$ 0.17$\pm$0.07 (5 Sy1s) versus 0.03$\pm$0.01 (4 Sy2s).

The reddening of the nuclear spectrum is on average higher
in the Seyfert 2s in our sample ($<$$E_{B - V}$$>_{\rm 6 Sy2s}$ $\sim$
0.55$\pm$0.07\,mag
versus $<$$E_{B - V}$$>_{\rm 6 Sy1s}$ $\sim$ 0.42$\pm$0.06\,mag). 
The electron densities are 
comparable in both objects ($<$$n_{e}$$>_{\rm 6 Sy2s} \sim$
1070$\pm$180\,cm$^{-3}$ versus $<$$n_{e}$$>_{\rm 6 Sy1s} \sim
1100\pm315$\,cm$^{-3}$). 
The ionisation parameter is on average higher in Seyfert-1 galaxies [$<$$U_{\rm log (n_e) = 3}$$>_{\rm 5
Sy1s} \sim (4.42$$\pm$1.2)$\cdot 10^{-3}$  versus $<$$U_{\rm log (n_e) = 3}$$>_{\rm 5 Sy2s}
\sim$ (2.66$\pm$0.2)$\cdot 10^{-3}$],  also when excluding the exceptional
high value of $U$ seen in Fairall\,51 [$<$$U_{\rm log (n_e) = 3}$$>_{\rm 4
Sy1s} \sim (3.21$$\pm$0.4)$\cdot 10^{-3}$], but the distributions overlap.

Moreover, the comparison 
shows that higher temperatures occur in type-1 objects ($<$$T_{e}$$>_{\rm 4 Sy1s}$  $\sim$
33590$\pm$7070\,K versus $<$$T_{e}$$>_{\rm 4 Sy2s}$ $\sim$ 14470$\pm$440\,K). 
Note that the difference in the flux ratio of
[\ion{O}{iii}]\,$\lambda$4363\,\AA/[\ion{O}{iii}]\,$\lambda$5007\,\AA~seen
between Seyfert-1 and Seyfert-2 galaxies has been
interpreted by \citet{ost78} as a 
difference in densities ($n_{\rm H} \sim 10^{6
  - 7}$\,cm$^{-3}$ for Seyfert-1 galaxies and 
$n_{\rm H} < 10^5$\,cm$^{-3}$ for Sy2s).
However, we interpret it as a difference in temperature  
in agreement
with the suggestions by \citet{hec79} and \citet{coh83} ($T_e > 20000$\,K for
Sy1s; $T_e \sim 10000$\,K for Sy2s).

Differences between the NLRs in Seyfert-1 and Seyfert-2 galaxies are known
from both imaging and spectroscopy and have been discussed by various authors 
on the basis of the unified model [e.g.~\citet{mul96b, schm98,
  nag01,sch03b}]. 
Statistics have shown that high-ionisation emission lines as
well as those with high critical densities
tend to be stronger in Seyfert-1 galaxies than in type 2s 
[e.g.~\citet{shu81, schm98, nag00}].
One explanation is that the highly-ionised gas clouds are located
close to the nucleus and can be hidden by the dust torus 
\citep{mur98a, mur98b, nag00}.
On the contrary, \citet{schm98} proposed that the NLR sizes in Seyfert-1
galaxies are (intrinsically) smaller than those of type 2s
(and not only due to projection effects): If the torus of 
Seyfert-1 galaxies is more
likely to be aligned with the galaxy plane (but has random orientations in
Sy2s) and the ionisation cone in type-1 AGNs is thus
perpendicular to the galaxy plane, there is a smaller number of ionisation-bounded clouds
in Seyfert-1 galaxies.
Based on a sample of 355 Seyfert galaxies, 
\citet{nag01} favour the first explanation.

Compared to the Seyfert-2 galaxies, the Seyfert 1s in our sample show
on average higher iron emission line fluxes such as [\ion{Fe}{vii}] and
[\ion{Fe}{x}] relative to H$\beta$, i.e.~high-ionisation lines (upper
ionisation potential 125\,eV and 262.1\,eV, respectively),
as well as higher [\ion{O}{iii}]\,$\lambda$4363\,\AA~intensities, a 
line with a rather low ionisation potential compared to these iron lines
(upper ionisation potential of 54.9\,eV) 
but high critical densities (3.3 $\cdot 10^7$\,cm$^{-3}$;
Table~\ref{lineratio1}),
in agreement with the results of \citet{nag01}.
Moreover, we find Seyfert-1 galaxies tending to have
higher electron temperatures and ionisation parameters in their nuclear
spectra. While the central electron densities are comparable 
taking into account the large scatter
of electron densities within the individual Seyfert 1 and Seyfert 2 galaxies, the
nuclear reddening is on average higher in the six Seyfert-2 galaxies.

The higher average central ionisation parameter is related to the observation
of stronger fluxes of high-ionisation lines in Seyfert-1 galaxies and can be
explained likewise: If the high-ionisation lines, along with
the Balmer and [\ion{O}{iii}] emission
lines, originate in gas clouds close to the BLR, they may be partly
hidden by the dust torus in Seyfert-2 galaxies.
Our observations of higher nuclear reddening in
Seyfert-2 galaxies argue in favour of this scenario proposed by \citet{nag01}.
It is reasonable to assume that lines with high ionisation potential arise
closer to the photoionising source, leading to a stratification of emission
lines. This is comparable to what has been found for 
the BLR using reverberation-mapping:
Different lines have different time lags with lines from high-ionised gas
responding earlier, showing that the ionisation structure is radially
stratified [e.g. \citet{pet93}]. 

The reason that we observe comparable nuclear densities in 
both type-1 and type-2
Seyferts may lie in the correction of the central electron temperature:
When comparing the measured electron densities directly, i.e.~not correcting
for the temperature, we get on average slightly lower densities for Seyfert-1
galaxies (which have the higher central temperatures):
$n_{e,{\rm ave, 6 Sy1s}} \sim
825\pm170$\,cm$^{-3}$ versus $n_{e,{\rm ave, 6 Sy2s}} \sim
950 \pm$ 160\,cm$^{-3}$. Taking into account the critical 
densities of the involved
forbidden emission lines, we cannot rule out that the temperature we measure corresponds
to a region closer to the centre than the electron density (if the density
increase towards the centre): While the critical
densities of the [\ion{O}{iii}] lines are high [$3.3 \cdot
10^7$\,cm$^{-3}$ for $\lambda$4363\,\AA, 
$7 \cdot 10^5$\,cm$^{-3}$ for $\lambda$5007\,\AA], they are significantly lower for the
[\ion{S}{ii}]\,$\lambda\lambda$6716,6731\,\AA~lines (1500-3900\,cm$^{-3}$).
It implies that while the [\ion{O}{iii}] lines are still emitted in a
dense central region with e.g.~$n_e \sim
10000$\,cm$^{-3}$, allowing us to
measure the temperature close to the nucleus, both [\ion{S}{ii}] lines are
collisionally de-excited. Thus, the flux we measure in these lines comes from
regions with lower densities further out along our line-of-sight. 

The galaxies of the present sample
underscore the so-called temperature problem
[e.g. \citet{sto96}]
as it is known in photoionisation modelling.
This generally refers to the problem that photoionisation
models underpredict the temperature in the NLR clouds,
as measured by the ratio [\ion{O}{iii}]\,$\lambda\lambda$4363/5007\,\AA.
Solutions include the reduction of oxygen (metal) abundances
(leading to increased heating) and/or the presence of dust within the
NLR clouds \citep{kom97}, or the presence of a significant
fraction of matter-bounded
clouds within the NLR \citep{bin96}.
The presence of an inner high density component to solve
the temperature problem was rejected  by \citet{kom97}
because such a component would strongly boost [\ion{O}{i}]\,$\lambda$6300\,\AA.
Indeed, inspecting the dependence of [\ion{O}{i}] on radius, for
our sample we do not find evidence for strongly increased [\ion{O}{i}]
emission in the core, or at a certain radius.
The three proposed solutions appear to be consistent with our
data, even though we do not directly measure the metal abundances
or the fraction of matter-bounded clouds.

\subsection{Black hole masses}
BH masses can be estimated using several methods.
First, we estimated BH masses
from the luminosity
at 5100\AA~using the empirical formula found by
\citet{pet04} (in Table~\ref{bh} denoted as $M_{\rm BH, Peterson\,et\,al.}$):
\begin{eqnarray*}
\log \frac{M_{\rm BH}}{10^8 M_{\odot}} = (-0.12\pm0.07) + (0.79\pm0.09) \cdot \log \frac{\lambda L_{\lambda} (5100 \AA)}{10^{44} 
\,{\rm erg\,s^{-1}}} \hspace{0.2cm} .
\end{eqnarray*}
To obtain 
$5100 \cdot L_{5100}$, we multiplied 5100\AA~by the monochromatic flux 
at the (redshifted) 5100\AA~continuum of the nuclear spectrum. 
We used the broad H$\alpha$ to H$\beta$ ratio
to correct for the reddening of the luminosity 
(except for MCG\,-6-30-15
for which we used the continuum reddening instead).

Another estimation of the central BH mass is obtained by first estimating
the radius of the BLR
\citep{kas05} (eq. 2):
\begin{eqnarray*}
\frac{R_{\rm BLR}}{10\,{\rm lt-days}} &=& (2.23\pm0.21) \cdot \left[\frac{\lambda L_{\lambda} (5100 \AA)}{10^{44} 
\,{\rm erg\,s^{-1}}}\right]^{0.69\pm0.05} \hspace{0.2cm} ,
\end{eqnarray*}
and then calculating the virial reverberation mass, correcting 
$v_{\rm FWHM, \,H\beta}$ by the empirical factor of $\sqrt{5.5}/2$
[derived by normalising the AGN $M_{\rm BH} - \sigma_{\star}$ relationship
to the $M_{\rm BH} - \sigma_{\star}$ relationship for quiescent galaxies,
\citet{pet04}]:
\begin{eqnarray*}
\frac{M_{\rm BH}}{10^8 M_{\odot}} &=& 0.02685 \cdot \frac{R_{\rm BLR}}{10\,{\rm lt-days}} \cdot \left(\frac{{v}_{\rm FWHM,\,H\beta}}{10^3\,{\rm km\,s^{-1}}}\right)^2
\end{eqnarray*}
(in Table~\ref{bh} denoted as $M_{\rm BH, Kaspi\,et\,al.}$).
The FWHM of the broad H$\beta$ emission line was determined
using only two fits to the observed H$\beta$ line, one for the narrow H$\beta$ component and one
for the broad one.
For  MCG\,-6-30-15, where we could not disentangle the broad
and the narrow H$\beta$ component, we used a fit to the total line profile
and thus, the BH mass estimation ($M_{\rm BH, Kaspi\,et\,al.}$)
can be considered as a lower limit (the line is dominated by the
broad component such an approach is reasonable).
Note that using the empirical factor of $\sqrt{5.5}/2$ instead of $\sqrt 3/2$ 
[as previously
assumed for an isotropic velocity dispersion and $\sigma_{\rm line}$
= FWHM/2; e.g. \citet{net90}] results in a $\sim$1.8
times higher BH mass \citep{pet04}.

Third, for comparison, we estimated BH masses from the 
 stellar velocity dispersion
[$\sigma_{\star}$; e.g. \citet{mer01}] using the  formula
from \citet{tre02}:
\begin{eqnarray*}
\log \frac{M_{\rm BH}}{M_{\odot}} = (8.13\pm0.06) + (4.02\pm0.32) \cdot \log \frac{\sigma_{\star}}{200\,{\rm km\,s^{-1}}} \hspace{0.2cm} .
\end{eqnarray*}
The stellar velocity dispersions for five of our six Seyfert-1 galaxies were derived
by taking the average of $\sigma_{\star}$'s obtained by \citet{gar05}
through two different methods (direct fitting and cross-correlation).
Note that their $\sigma_{\star}$ was measured within
the aperture of $\sim$2\arcsec$\times$2\farcs5. However, we did not correct for the aperture
size [see \citet{tre02} for an extensive discussion
on this topic].

The results are summarised in Table~\ref{bh}. The difference
in the derived BH masses for the first two methods can be as
high as a factor of 3 which is in the range of the 3$\sigma$ error.
$M_{\rm BH, Tremaine\,et\,al., \sigma_{\star}}$ and 
$M_{\rm BH, Peterson\,et\,al.}$ are in agreement to within 1$\sigma$
for most galaxies. 
For NGC\,526a, 
$M_{\rm BH, Tremaine\,et\,al., \sigma_{\star}}$ is significantly larger
(by a factor $\sim$8) than $M_{\rm BH, Peterson\,et\,al.}$. 
On the other hand, 
for MCG\,-6-30-15, the 
$M_{\rm BH, Tremaine\,et\,al., \sigma_{\star}}$ is by a factor of 3 smaller than
$M_{\rm BH, Peterson\,et\,al.}$ but in agreement with $M_{\rm BH, Kaspi\,et\,al.}$
within the errors.
However, all these values were derived from statistical formulae
from which individual galaxies may deviate quite a lot.

We searched the literature for other BH mass estimations for our
target galaxies but were successful only for MCG\,-6-30-15 for which
the results agree within the errors (see Appendix~\ref{mcg6}).

\begin{table*} 
\begin{minipage}{180mm}
\caption[]{\label{bh} BH masses\footnote{The errors given for the BH masses reflect the errors of our measurements of $L_{5100}$ and FWHM$_{\rm H\beta}$, the error in $\sigma_{\star}$ as well as the statistical errors of the three different fitting relations (Peterson et al., Kaspi et al., Tremaine et al.).}}
\begin{center}
\begin{tabular}{lcccccccc}
\\[-2.3ex]
\hline
\hline\\[-2.3ex]
\multicolumn{1}{c}{Galaxy} & $5100 \cdot L_{5100}$ & FWHM$_{\rm H\beta}$\footnote{Broad H$\beta$ of 2 component fit. The FWHMs of the corresponding narrow H$\beta$ component of the five galaxies (excluding MCG\,-6-30-15)
in the order of listing are: 444, 480, 427, 406, and 445\,km\,s$^{-1}$, respectively.}
& $M_{\rm BH, Peterson\,et\,al.}$ & $M_{\rm BH, Kaspi\,et\,al.}$ & $M_{\rm BH, Tremaine\,et\,al., \sigma_{\star}}$\\[0.25ex]
&  (10$^{44}$ erg\,s$^{-1}$) &  (km\,s$^{-1}$) & (10$^8$ $M_{\odot}$) &  (10$^8$ $M_{\odot}$) &  (10$^8$ $M_{\odot}$)\\[0.25ex]
\hline\\[-2.3ex] 
Fairall\,51 & 1.53$\pm$0.2  & 3330$\pm$300 & 1.1 (+0.4-0.3) & 0.9 (+0.4-0.3) & --\footnote{No $\sigma_{\star}$ measurement available in literature}\\
NGC\,6860 & 0.44$\pm$0.05  & 5920$\pm$600 & 0.4$\pm$0.1 & 1.2 (+0.6-0.4)
& 0.4 (+0.5-0.2)\\
Mrk\,915 & 1.68$\pm$0.2  & 4560$\pm$500 & 1.1 (+0.4-0.3) & 1.8 (+0.9-0.6) & 0.6 (+0.9-0.4)\\
NGC\,526a & 0.17$\pm$0.02  & --\footnote{No broad H$\beta$ line} & 0.19 (+0.09-0.07) & --$^d$ & 1.6 (+1.2-0.8)\\
MCG\,-05-13-017 & 0.41$\pm$0.04  & 5240$\pm$500 & 0.37 (+0.13-0.11) & 0.9 (+0.4-0.3) & 0.24 (+0.15-0.11)\\[0.1ex]
MCG\,-6-30-15 & 0.27$\pm$0.03
& 1990$\pm$200 & 0.27 (+0.11-0.09) & 0.1 (+0.05-0.04)\footnote{This is a lower limit for the BH mass since the narrow H$\beta$ component was not removed.} & 0.08 (+0.07-0.05)\\[0.1ex]
\hline\\[-2.3ex]
\end{tabular}
\end{center}
\end{minipage}
\end{table*}

\subsection{Reddening distribution}
The reddening was derived from the
recombination value of the narrow H$\alpha$/H$\beta$ emission-line ratio
except for MCG\,-6-30-15 where we
could not disentangle the broad and narrow Balmer lines (in the central
$\sim$3\arcsec). Instead, we show for this
object the 
reddening distribution of the continuum with respect to the
stellar template (see also paper I \& II). 
On the contrary, the continuum slope reddening is not available for the other type 1s
as no stellar template was fit. 
While the nuclear reddening is given in Table~\ref{result}, 
we give in Table~\ref{reddening} the highest reddening value within the NLR, 
the distance from the centre at which it occurs as well as the global 
reddening, 
 i.e.~derived
from the total H$\alpha$ and H$\beta$ fluxes within the NLR.

While the highest reddening value within the NLR is on average
slightly higher in Seyfert 2s (0.75$\pm$0.06) than in Seyfert 1s (0.57$\pm$0.07),
we find that the reddening derived from the global Balmer decrement
is comparable in Seyfert 1s and 2s (see Table~\ref{reddening} and paper II, Table 6):
$<$$E_{B - V}$$>_{\rm 5 Sy1s}$ $\sim$ 0.37$\pm$0.04\,mag
(excluding MCG-6-30-15) and
$<$$E_{B - V}$$>_{\rm 6 Sy2s}$ $\sim$ 0.40$\pm$0.04\,mag.
When excluding NGC\,526a which can
be considered as a galaxy of transient Seyfert-type 1.9,
the average reddening value for four Seyfert-1 galaxies is indeed the
same as that for Sy2s:
$<$$E_{B - V}$$>_{\rm 4 Sy1s}$ $\sim$ 0.41$\pm$0.03\,mag.

This finding is opposite to the results of \citet{rhe05} who concluded
that Sy 1s have much lower (or zero) reddening than Sy 2s, based on near-IR line ratios.
They speculate that the difference could be caused either by
a large-scale ($>$100pc) torus or by an intrinsically different
grain size distributions in Sy 1s and 2s.
Our values rather agree with previous measurements
[e.g. \citet{coh83, gas84, tsv89}]: Although
these authors find slightly larger values of reddening in Sy2s,
substantial reddening is present in Sy 1s as well.

\begin{figure*}
\includegraphics[width=18cm]{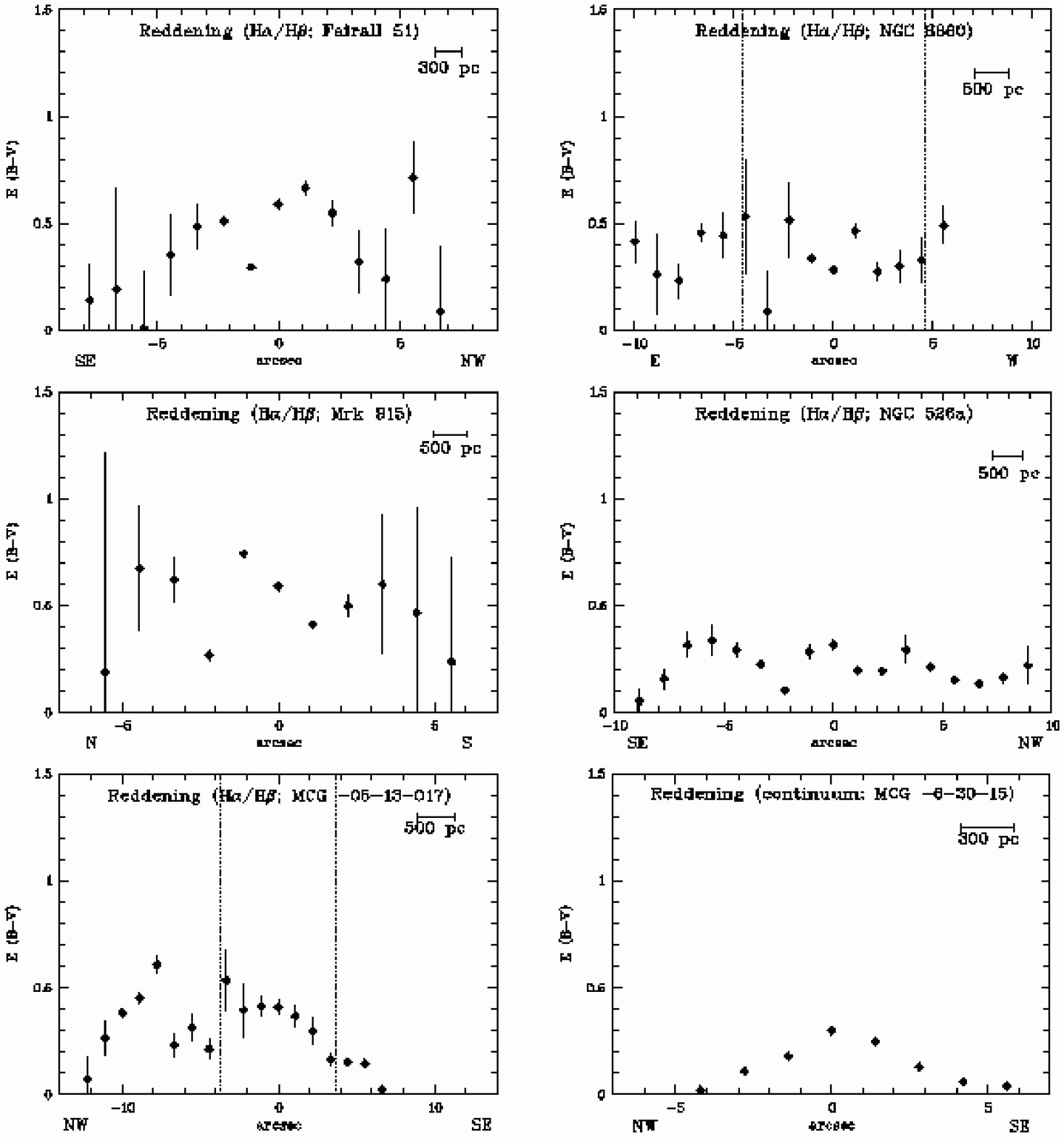}
\caption[]{\label{reddening1} \small
Reddening distributions of 
Fairall\,51,
  NGC\,6860, Mrk\,915, NGC\,526a, MCG\,-05-13-017, and MCG\,-6-30-15.
The edge of the NLR as determined from the
  diagnostic diagrams is indicated by dotted lines (NGC\,6860 and MCG\,-05-13-017).}
\end{figure*}

In Fig.~\ref{reddening1}, we show radial profiles of the reddening (for Sy 2s, see
Fig.~5 in paper I and Fig.~3 in paper II).
Among Sy1s, there are clear spatial gradients of the reddening
with  $E_{B - V}$ peaking at or near the photometric centre
in Mrk\,915 and MCG\,-6-30-15 (note that the latter was
determined from the continuum); in other Sy 1s, the reddening is more
even or patchy within the NLR.
Among Sy 2s, the reddening clearly peaks at or near the photometric
centre in IC\,5063, ESO\,362-G008 and NGC\,5643; there are also
systematic spatial gradients in NGC\,1386, NGC\,7212, 
and NGC\,3281, though the
maximum reddening does not coincide with the photometric centre.

In the (online) appendix, we give
the reddening of the BLR derived from the broad Balmer decrement
(when discussing the objects individually).

\begin{table}
\begin{minipage}{80mm}
 \caption[]
{\label{reddening} Maximum and global reddening within the NLR\footnote{We excluded MCG\,-6-30-15 as we do not have a measure of the reddening from the H$\alpha$ to H$\beta$ ratio.}}
\begin{center}
\begin{tabular}{lccc}
\\[-2.3ex]
\hline
\hline\\[-2.3ex]
\multicolumn{1}{c}{Galaxy} & max. $E_{B - V}$\footnote{Highest reddening value within the NLR} & Distance\footnote{Distance from the centre of highest reddening value} & global $E_{B - V}$\footnote{Derived by adding the H$\alpha$ and H$\beta$ flux within the NLR}\\
& (mag) & (\arcsec) & (mag) \\[0.25ex]
\hline\\[-2.3ex]
Fairall\,51     & 0.72$\pm$0.2 & 5.55 & 0.39$\pm$0.06\\
NGC\,6860       & 0.53$\pm$0.3 & -4.44 & 0.36$\pm$0.04\\
Mrk\,915        & 0.74$\pm$0.02 & -1.11 & 0.50$\pm$0.05\\
NGC\,526a       & 0.34$\pm$0.07 & -5.55 & 0.22$\pm$0.02\\
MCG\,-05-13-017 & 0.54$\pm$0.1 & -3.33 & 0.38$\pm$0.04\\[0.1ex]
\hline\\[-2.3ex]
\end{tabular}
\end{center}
\end{minipage}
\end{table}

\subsection{Spatially resolved spectral diagnostics}
\label{2ddiag}
In paper I \& II, we described the use of 
diagnostic line-ratio diagrams
of the three types pioneered by \citet{bal81}
to not only distinguish
bet\-ween emission-line object classes (e.g. Seyfert galaxies, LINERs, starbursts,
transition objects), but to  probe the ``real'' NLR size, i.e. the central region which
is photoionised by the AGN, and to discriminate the contribution from starbursts.
Such an approach has already been chosen by other authors to study
the ionisation mechanism in the circumnuclear and extranuclear regions of Seyfert galaxies
[e.g.~\citet{rad98, tem03, cir05}]. It often reveals that emission-line ratios
at larger distances from the central AGN change 
towards \ion{H}{ii} region-like ones due to an increasing contribution
to the ionisation by surrounding star-forming regions.

\begin{figure*}
\includegraphics[width=18cm]{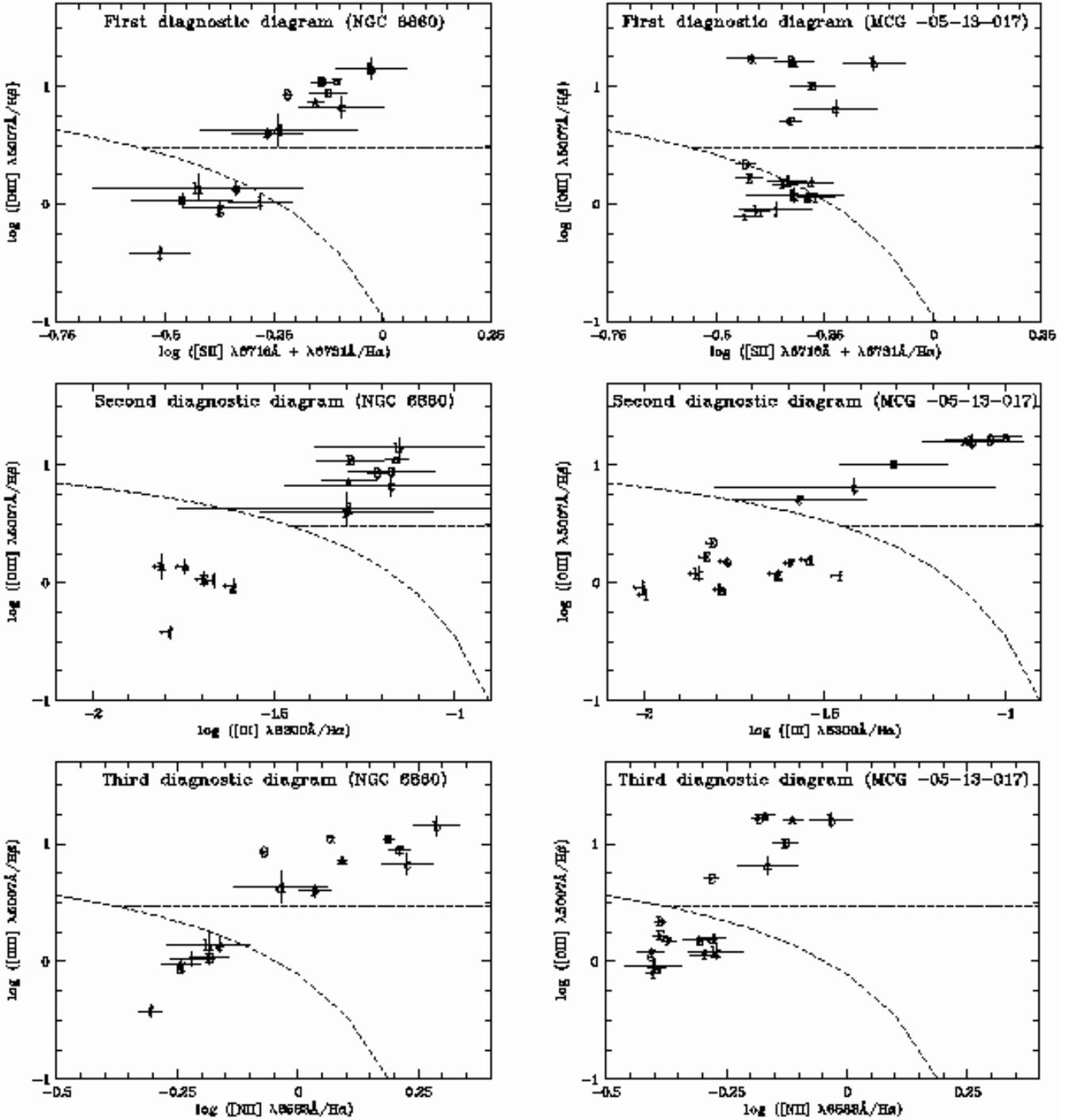}
\caption[]
{\label{diag1} \small
All three diagnostic diagrams for
spatially-resolved emission-line ratios in 
NGC\,6860 ({\it left panels}) and MCG\,-05-13-017 
({\it right panels}).}
\end{figure*}

\begin{figure*}
\includegraphics[width=18cm]{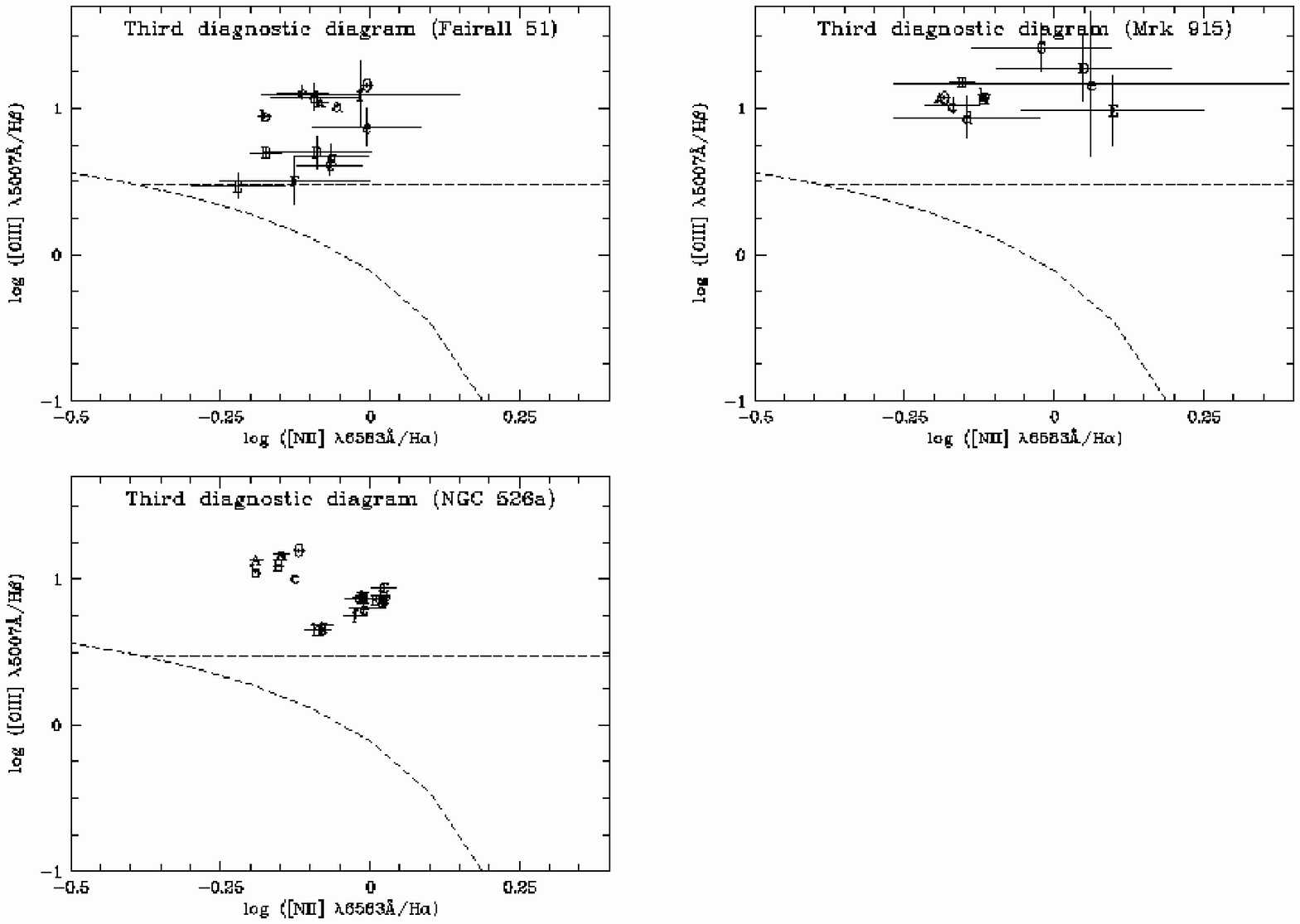}
\caption[]
{\label{diag3} \small
Emission-line ratios in the third diagnostic diagram for Fairall\,51,
Mrk\,915, and NGC\,526a.
All line ratios fall in the AGN regime. 
}
\end{figure*}

The high S/N ratio of our spectra enables us 
to measure line ratios for all three diagrams (``first'': [\ion{O}{iii}]/H$\beta$ versus [\ion{S}{ii}]/H$\alpha$;
``second'': [\ion{O}{iii}]/H$\beta$ versus [\ion{O}{i}]/H$\alpha$; ``third'': 
[\ion{O}{iii}]/H$\beta$ versus [\ion{N}{ii}]/H$\alpha$) out to several arcseconds from the
nucleus (Figs.~\ref{diag1},~\ref{diag3}).
The symbols are chosen such that ``O'' refers to
the central spectrum, the small letters mark regions corresponding to ``-''
arcseconds from the nucleus, the capital
ones mark regions corresponding to ``+''
arcseconds from the nucleus (Table~\ref{tablediag}).
In the second diagnostic diagram, the data
points of the outer regions are upper limits, due to the
faintness of the [\ion{O}{i}]\,$\lambda$6300\,\AA~line involved.

As for NGC\,1386 and NGC\,5643 (paper I \& II),
we find a clear transition between line ratios falling in the AGN
regime and those typical for \ion{H}{ii} regions
in two Seyfert-1 galaxies of our sample (NGC\,6860 and MCG\,-05-13-017).
We present all three diagnostic diagrams of these objects
in Fig.~\ref{diag1}.

For the remaining four galaxies, no such transition is observed but all
emission-line ratios  are typical for gas 
ionised by an AGN power-law continuum.
As the distributions in the three diagnostic diagrams are comparable, we
present only the third diagnostic diagram for these objects in
Fig.~\ref{diag3}. (We do not show the diagnostic diagram for
MCG\,-6-30-15 as we could not disentangle the broad and narrow Balmer
emission lines in the central $\sim$3\arcsec.)

We use the diagnostic diagrams to determine the NLR size. The results are
summarised in Table~\ref{tablediag}. For those objects which show a transition
of emission-line ratios from the central AGN region to \ion{H}{ii} regions,
this method gives a measure of the NLR size without [\ion{O}{iii}] contamination from circumnuclear starbursts:
Although \ion{H}{ii} regions may be
present over the entire emission-line region, 
the AGN ionisation dominates in the innermost arcseconds,
determining the size of the NLR.

For both objects with such a transition, 
the determined NLR size is about twice as large
as that measured from the HST snapshot survey of \citet{sch03a}, 
showing the low sensitivity of this survey.
On the other hand, some authors have attributed all [\ion{O}{iii}] emission
to the extended NLR:
For MCG\,-05-13-017, \citet{fra00} give a size of $\sim$17\arcsec~for the extended NLR,
while our diagnostic diagrams reveal that only the central $\pm$3\arcsec~consist of
gas ionised by the central AGN. From emission line ratios, \citet{lip93} classify 
NGC\,6860 as transitional object between Seyfert galaxies and starbursts.
However, we can show that NGC\,6860 is a Seyfert galaxy with the
NLR extending out to $r \sim 5$\arcsec~and surrounding starbursts, giving rise to
[\ion{O}{iii}] emission out $r \sim 10$\arcsec.

To conclude, compared to the spatially resolved
spectral diagnostics measuring the ``real'' NLR size, the apparent NLR size determined
by [\ion{O}{iii}] images can be either smaller in case of low sensitivity 
or larger in case of contributions of circumnuclear starbursts. 
For the remaining four objects, the estimated NLR size is a lower limit,
pointing out the limitations of this method (see paper II for discussion).

In paper I, we used \texttt{CLOUDY} photoionisation modelling
to show  that the observed distinction between \ion{H}{ii}-like and
AGN-like line ratios represents a true difference in
ionisation source, and that our method to measure the NLR radius
is valid. These results can also be applied here.
The second diagnostic diagram including 
the [\ion{O}{i}] emission-line is essential to reach
this conclusion, since  
our photoionisation calculations showed that a combination
of outwards decreasing ionisation parameter and metal abundances
could mimic \ion{H}{ii}-like line ratios despite
an intrinsic AGN ionisation source in the [\ion{O}{iii}]/H$\beta$ versus [\ion{N}{ii}]/H$\alpha$ and 
the [\ion{O}{iii}]/H$\beta$ versus [\ion{S}{ii}]/H$\alpha$ diagrams.

\begin{table*}
\begin{minipage}{180mm}
 \caption[]
{\label{tablediag} Results from diagnostic diagrams\footnote{The second column gives 
the distance from the centre to the first spectra
(marked with the letters ``a'' and ``A'' in the diagnostic diagrams). 
In the third column, 
the orientation of the small and capital letters is listed.
The maximum [\ion{O}{iii}] radius (S/N $>$ 3) at the
same p.a.~taken from literature is given in the fourth column.
We also give the
 [\ion{O}{iii}] radius (S/N $>$ 3) observed from our
spectra (column 5). In the sixth column, the radius is given until which we
were able to plot line ratios in the diagnostic diagrams.  In the last column, 
the radius of
the NLR as determined from the diagnostic diagrams is given in \arcsec~and, in
brackets, pc, respectively.
The two objects with a clear transition between
NLR and \ion{H}{ii} region are marked in bold.}}
\begin{center}
\begin{tabular}{lcr@{/}lcccc}
\\[-2.3ex]
\hline
\hline\\[-2.3ex]
\multicolumn{1}{c}{Galaxy} & ``a/A'' & \multicolumn{2}{c}{``a/A''} &  $R_{\rm [OIII]}$ & $R_{\rm [OIII]}$ & $R_{\rm line-ratios}$ & $R_{\rm NLR}$\\
& Distance (\arcsec) & \multicolumn{2}{c}{Orientation} & Literature (\arcsec) & Our Data (\arcsec) & Our data (\arcsec) & Our Data (\arcsec, pc)\\[0.25ex]
\hline\\[-2.3ex]
Fairall\,51              & 1   & SE & NW & 2\footnote{ Taken from HST image of \citet{sch03a}}              & \hspace{+1.5mm}9   & \hspace{+1.5mm}8          & $>$8 (2260)\\
{\bf NGC\,6860}          & 1   & E & W   & 3$^b$               & 10             & 10 & \hspace{+1.5mm}{\bf 5 (1465)}\\
Mrk\,915                 & 1   & N & S   & 2$^b$               & 12             & \hspace{+1.5mm}6 & $>$6 (2720)\\
NGC\,526a                & 1   & SE & NW & \hspace{-2mm}11\footnote{Taken from groundbased image of \citet{mul96a}} & 20  & \hspace{+1.5mm}9           & $>$9 (3260)\\
{\bf MCG\,-05-13-017} & 1   & NW & SE & 1$^b$               & 17 & 11 & \hspace{+0.0mm}{\bf 3 (730)}\\
MCG\,-6-30-15         & 1.4 & NW & SE & 2$^b$               & 12  & \hspace{+1.5mm}4  & \hspace{+2.mm}4? (700)\footnote{In the central 3\arcsec~of MCG\,-6-30-15, we cannot disentangle the broad
and narrow Balmer components and therefore do not determine the line
ratios. In the outer region to a distance of $\pm$4\arcsec, they fall in the
AGN regime.}\\[0.1ex]
\hline\\[-2.3ex]
\end{tabular}
\end{center}
\end{minipage}
\end{table*}

\subsection{Surface-brightness distribution}
\label{longsur}
\begin{figure*}
\includegraphics[width=18cm]{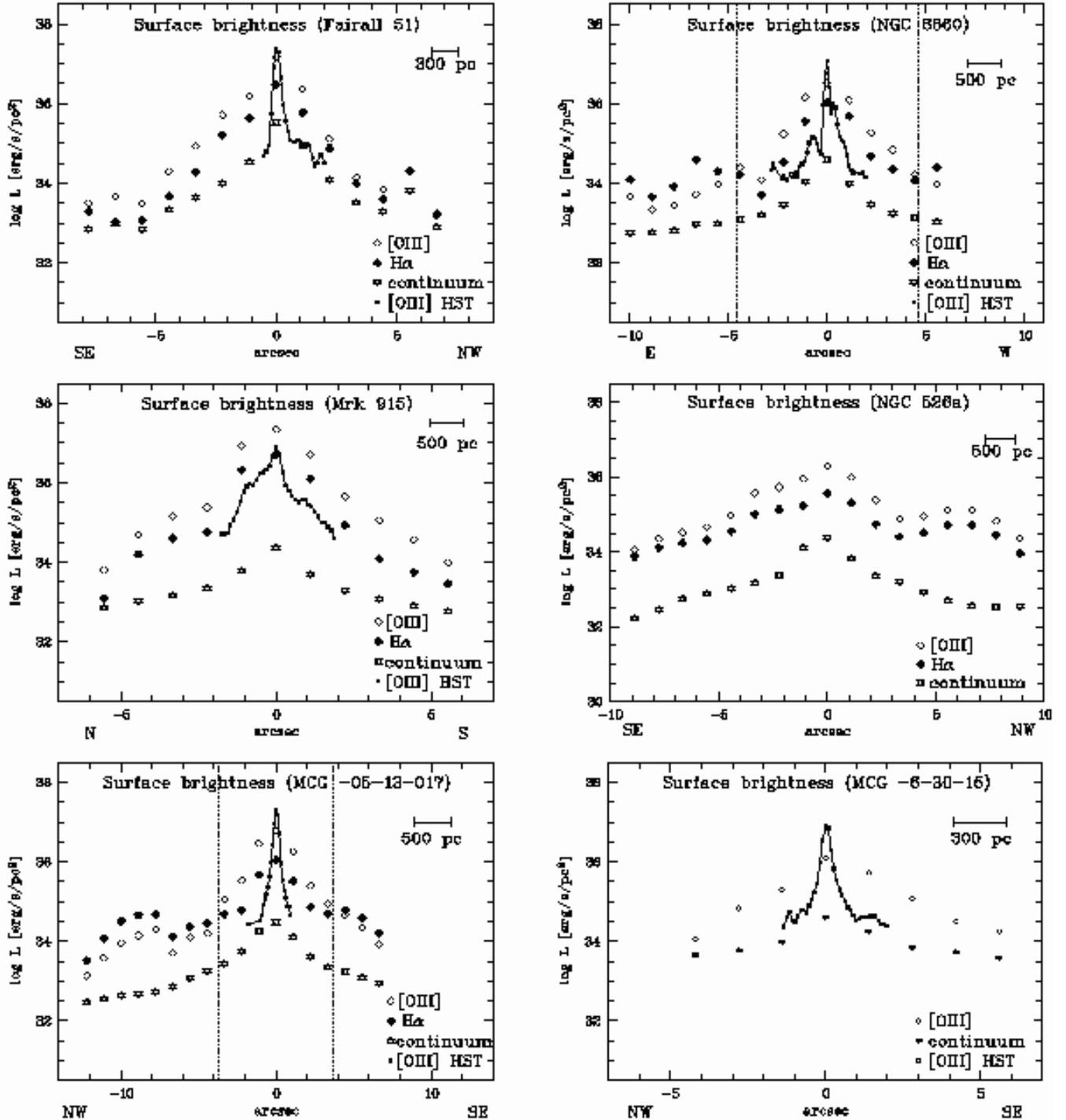}
\caption[]
{\label{lum1} \small
Surface-brightness distributions of 
Fairall\,51, NGC\,6860, Mrk\,915, NGC\,526a, MCG\,-05-13-017,
and MCG\,-6-30-15
in [\ion{O}{iii}] (open diamonds), 
narrow H$\alpha$ (filled diamonds), and continuum (at
  5450-5700\,\AA, stars). The [\ion{O}{iii}] surface-brightness distribution
  from the HST image is shown as small open squares connected by a line
(HST pixel scale $\sim$ 0\farcs1\,pix$^{-1}$). 
Only data points with S/N $>$ 3
 were included. 
Error bars are smaller than the symbol size.
The HST image has a 10 to 20 times higher spatial resolution but
a significantly lower sensitivity, not allowing to measure the outer parts
of the NLR. 
The edge of the NLR as determined from the
  diagnostic diagrams is indicated by dotted lines (NGC\,6860 and MCG\,-05-13-017).
Note that NGC\,526a is not included in the HST snap-shot survey by \citet{sch03a}.
}
\end{figure*}

The spatially varying luminosities in the [\ion{O}{iii}] and (narrow) H$\alpha$ emission lines as well as
the continuum (at 5450-5700\,\AA) were calculated and divided by the corresponding area in square
  parsecs at the galaxy to allow a comparison among all galaxies in our sample
(Fig.~\ref{lum1}).
The surface-brightness distributions are similar to each other, centrally peaked and 
decreasing with distance from the nucleus. 

For comparison, the [\ion{O}{iii}] surface-brightness distributions
from the HST images of \citet{sch03a} are shown for those objects included in the
  HST snapshot survey. 
They were derived by averaging three vectorplots along the major axis of the NLR
emission (see also paper I \& II). In all objects, they clearly show the
higher spatial resolution of the HST image (0\farcs05 - 0\farcs1
pix$^{-1}$) compared to the 1-2\arcsec~spatial sampling of our spectral data.
However, they also reveal the low sensitivity of the HST images
compared to our spectroscopy: The
[\ion{O}{iii}] emission at a S/N of 3 ends significantly earlier than what can
be seen in our spectral data.
In some cases, the HST [\ion{O}{iii}] surface-brightness distributions reveal several
subpeaks of possibly individual NLR clouds, as can be already seen in the
[\ion{O}{iii}] images (Fig.~\ref{galaxies1}).
These substructures are smoothed out in our $\sim$10-20 times lower spatial resolution
spectra but are nevertheless still visible as a secondary or tertiary peak,
mostly in emission lines.

We fitted a power-law function $L = L_{0} (\frac{R}{R_0})^{\delta}$ (with projected radius $R$) to the
surface-brightness distributions of [\ion{O}{iii}], H$\alpha$, and the continuum. 
The fitting parameters are shown in Table~\ref{fitlum} 
(with $L_0$ referring to $R_0$ = 100 pc
from the nucleus). 
Only data points within the NLR were included and the central point was excluded from the fit.

The [\ion{O}{iii}] surface brightness falls faster with radius than the
H$\alpha$ surface brightness and also faster than the continuum
($<$$\delta_{\rm [OIII]}$$> \sim -2.95\pm0.3$; 
$<$$\delta_{\rm H\alpha}$$> \sim -2.58\pm0.5$;
$<$$\delta_{\rm cont}$$> \sim -1.63\pm0.2$).
The average slope for both the [\ion{O}{iii}] and H$\alpha$ surface
brightness gets even steeper when excluding NGC\,526a which can
be considered as a galaxy of transient Seyfert-type 1.9
($<$$\delta_{\rm [OIII]}$$> \sim -3.19\pm0.2$; 
$<$$\delta_{\rm H\alpha}$$> \sim -2.9\pm0.4$).

For all three surface-brightness distributions ([\ion{O}{iii}], H$\alpha$, continuum),
Seyfert-1 galaxies show a steeper radial slope than Seyfert 2s (see paper II)
($<$$\delta_{\rm [OIII]}$$>_{\rm 6 Sy1} \sim -2.95\pm0.3$ versus
$<$$\delta_{\rm [OIII]}$$>_{\rm 5 Sy2} \sim -2.24\pm0.2$;
$<$$\delta_{\rm H\alpha}$$>_{\rm  5 Sy1} \sim -2.58\pm0.5$ versus
$<$$\delta_{\rm H\alpha}$$>_{\rm  5 Sy2} \sim -2.16\pm0.2$; 
$<$$\delta_{\rm cont}$$>_{\rm  6 Sy1} \sim -1.63\pm0.2$ versus
$<$$\delta_{\rm cont}$$>_{\rm  5 Sy2} \sim -1.19\pm0.1$), 
a difference that is even more pronounced when excluding NGC\,526a (see above).

We want to point out that the continuum slope for the Seyfert-1 galaxies may
be boosted by the AGN as we only excluded the nuclear datapoint
but no other datapoints within the seeing
range (1-2\arcsec) which may still be contaminated by the unresolved AGN contribution;
excluding these datapoints leaves us with too few datapoints in most cases.
However, to estimate this effect, we calculated the average continuum slope excluding the central 2 arcseconds
for four Seyfert-1 galaxies for which 3-7 datapoints remain in the fit.
It is still steeper than that for the Seyfert-2 galaxies:
$<$$\delta_{\rm cont}$$>_{\rm  4 Sy1} \sim -1.58\pm0.2$ versus
$<$$\delta_{\rm cont}$$>_{\rm  5 Sy2} \sim -1.19\pm0.1$.

\begin{table*} 
\begin{minipage}{180mm}
\caption[]{\label{fitlum} Fitting parameters of surface-brightness distributions\footnote{A linear least-squares fit was applied with $\log L = \delta \cdot
\log R/R_0 + \log L_{0}$. $L_0$ corresponds to $R_0$ = 100 pc
from the nucleus.
The number of data points included in the fit is
given in column 2 (= half the number of averaged values from both sides of
the nucleus). For those objects which show a transition between line
ratios typical for AGNs and \ion{H}{ii}-region like ones in the diagnostic
diagrams, determining the size of the NLR, only data points within the NLR
were included (NGC\,6860 and MCG\,-05-13-017). 
For
MCG\,-6-30-15, no deconvolution of the broad and narrow H$\alpha$ was possible.}}
\begin{center}
\begin{tabular}{lccccccc}
\\[-2.3ex]
\hline
\hline\\[-2.3ex]
\multicolumn{1}{c}{Galaxy} & Data Points & \hspace{+3mm}$\delta_{\rm [OIII]}$ & $\log L_{\rm [OIII], 0}$ & \hspace{+3mm}$\delta_{\rm H\alpha}$ & $\log L_{\rm H\alpha, 0}$ & \hspace{+3mm}$\delta_{\rm cont}$ & $\log L_{\rm cont, 0}$\\
& & & (erg\,s$^{-1}$\,pc$^{-2}$) & & (erg\,s$^{-2}$\,pc$^{-2}$) & & (erg\,s$^{-2}$\,pc$^{-2}$)\\[0.25ex]
\hline\\[-2.3ex] 
Fairall\,51        & 6               & $-$3.55$\pm$0.25 & 38.14 & $-$3.16$\pm$0.48 & 37.36 & $-$2.15$\pm$0.30 & 35.79 \\
NGC\,6860          & 4               & $-$3.06$\pm$0.12 & 37.69 & $-$2.59$\pm$0.45 & 36.85 & $-$1.62$\pm$0.43 & 36.27 \\
Mrk\,915           & 5               & $-$3.92$\pm$0.32 & 39.58 & $-$3.88$\pm$0.33 & 38.89 & $-$1.72$\pm$0.45 & 35.74 \\
NGC\,526a          & 8               & $-$1.72$\pm$0.19 & 37.1 & $-$1.28$\pm$0.19 & 36.13 & $-$1.71$\pm$0.06 & 34.99 \\
MCG\,-05-13-017 & 3               & $-$2.90$\pm$0.07 & 37.61 & $-$1.98$\pm$0.48 & 36.4 & $-$1.66$\pm$0.02 & 34.9 \\[0.1ex]
MCG\,-6-30-15   & 3               & $-$2.52$\pm$0.41 & 36.58 & \hspace{+3mm}--   & --   & $-$0.94$\pm$0.10 & 34.49 \\[0.1ex]
\hline\\[-2.3ex]
\end{tabular}
\end{center}
\end{minipage}
\end{table*}

\subsection{Electron-density distribution}
\label{longdens}
Applying the classical methods outlined in \citet{ost89},
we derive the electron density as a function of distance
to the nucleus using the ratio of the 
[\ion{S}{ii}]\,$\lambda$$\lambda$6716,6731\,\AA~pair 
of emission lines.
We used the observed central temperature to 
correct for the dependency of electron
density on temperature\footnote{$n_e ({\rm T}) 
= n_e ({\rm [SII]\,ratio}) \cdot \sqrt{(T/10000)}$}. 
Due to the faintness of the involved
[\ion{O}{iii}]\,$\lambda$4363\,\AA~emission line, we were not able to measure
the temperature in the outer parts.
For those objects for which no temperature was determined, we assumed  $T =
10000$\,K.

\begin{figure*}
\includegraphics[width=18cm]{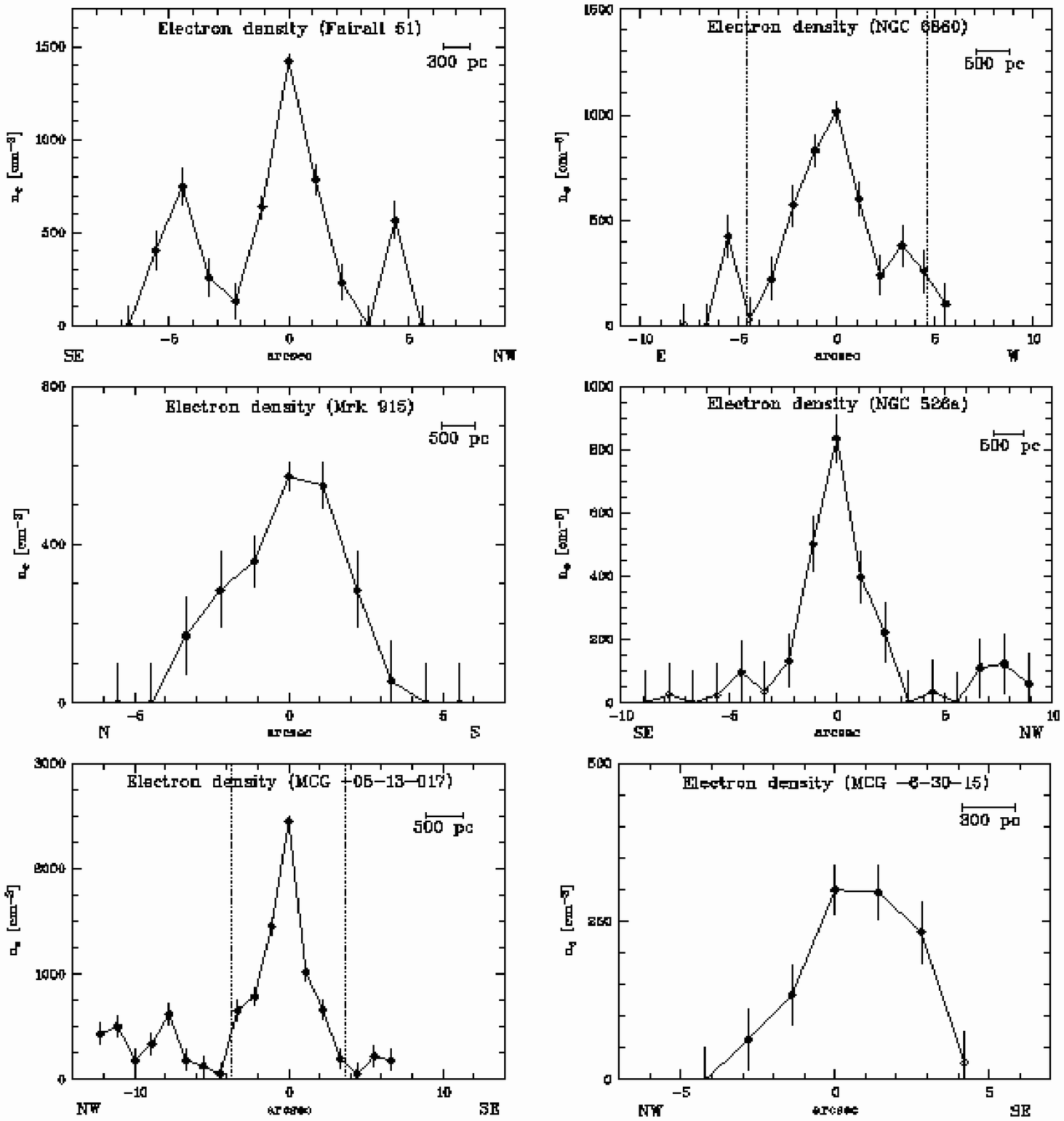}
\caption[]
{\label{density1} \small
Electron density obtained
from the [\ion{S}{ii}]\,$\lambda$6716\,\AA/$\lambda$6731\,\AA~ratio 
as a function of the distance from the nucleus 
Fairall\,51, NGC\,6860, Mrk\,915, NGC\,526a, MCG\,-05-13-017,
and MCG\,-6-30-15. 
Open symbols indicate locations
where $n_{\rm e, obs}$ is in the low-density limit (assumed $\le$ 50\,cm$^{-3}$).
The edge of the NLR as determined from the
  diagnostic diagrams is indicated by dotted lines 
(NGC\,6860 and MCG\,-05-13-017).}
\end{figure*}

In all objects, 
the electron density is highest at the nucleus and 
decreases outwards down to the low-density limit
(assumed to be 50\,cm$^{-3}$; Fig.~\ref{density1}). In
some cases, 
it reveals a secondary or tertiary peak on one or both sides of the optical centre.
A characteristic structure with a central peak and a smaller peak on both
sides of the nucleus can be identified in four objects (Fairall\,51, NGC\,6860, NGC\,526a, MCG\,-05-13-017). The outer
peaks are often close to the boundary of the NLR. These density enhancements
may indicate shocks occurring at the edge of the NLR.

In Table~\ref{fitden}, we give the results of fitting a power-law function $n_{\rm e, obs}$
= $n_{e, 0}$$ (\frac{R}{R_0})^{\delta}$ to the observed electron densities
(with $n_{\rm e, 0}$ at $R_0$ = 100 pc
from the nucleus).
Note that we did include only data points within the NLR.
 $\delta$ ranges
between -0.9 and -2.3. On average, the density decreases with $R^{-1.46 \pm
  0.2}$.
Thus, Seyfert-1 galaxies tend to show a steeper slope
than Seyfert-2 galaxies ($<$$\delta$$>_{\rm 5 Sy2} \sim -1.14\pm0.1$; paper II).
However, the individual scatter is rather large.

The temperature can be a function of distance from the central
AGN. Unfortunately, we are not able to determine the temperature dependency on
distance from the nucleus. In those objects where we are able to trace the
electron temperature in the inner few arcseconds, it remains roughly constant.
One may expect that the temperature is decreasing if the AGN is the only heating source.
In that case, correcting with the central temperature overestimates
the electron density in the outer parts. The observed decreasing slope can
therefore not be artificially introduced by a wrong temperature correction.
On the other hand, some authors report an increasing temperature with
distance from the nucleus [e.g.~\citet{ber83}] and explain it with a decrease
in electron density faster than $n_e \propto r^{-2}$. However, the average decrease of
electron density $n_{\rm e, obs}$ 
we observe is with $\delta \sim -1.5$ slower than that.

Note that the critical density for
[\ion{S}{ii}]\,$\lambda\lambda$6716,6731\,\AA~is $\sim$1500\,cm$^{-3}$ and
3900\,cm$^{-3}$, respectively. 
Thus, these lines can only be used to measure the density in an environment
with densities below $\sim$1500\,cm$^{-3}$. At least for some objects
in which we measure central densities in this regime 
the central density may thus be underestimated.

\begin{table}
\begin{minipage}{80mm}
 \caption[]
{\label{fitden} Fitting parameters of electron-density distribution\footnote{A linear least-squares fit was applied with $\log$$n_{\rm e, obs} = \delta \cdot
\log R/R_0 + \log $$n_{e, 0}$.
$n_{e, 0}$ corresponds to the value at $R_0$ = 100 pc distance
from the centre.
The number of data points included in the fit is
given in column 2 (= half the number of averaged values from both sides of
the nucleus). For those objects which show a transition between line
ratios typical for AGNs and \ion{H}{ii}-region like ones in the diagnostic
diagrams, determining the size of the NLR, only data points within the NLR
were included (NGC\,6860, MCG\,-05-13-017).}}
\begin{center}
\begin{tabular}{lccc}
\\[-2.3ex]
\hline
\hline\\[-2.3ex]
\multicolumn{1}{c}{Galaxy} & Data Points & $\delta$ & $\log n_{e, 0}$ (cm$^{-3}$)\\[0.25ex]
\hline\\[-2.3ex]
Fairall\,51 &  6     & -2.10$\pm$1.50  & 4.1 \\
NGC\,6860 & 4        & -1.06$\pm$0.22 & 3.4 \\
Mrk\,915 &  3        & -1.20$\pm$0.40  & 3.5 \\
NGC\,526a & 8        & -1.15$\pm$0.50  & 3.1 \\
MCG\,-05-13-017 & 3 & -0.94$\pm$0.14 & 3.5\\[0.1ex]
MCG\,-6-30-15 & 3 & -2.32$\pm$1.42 & 3.4\\[0.1ex]
\hline\\[-2.3ex]
\end{tabular}
\end{center}
\end{minipage}
\end{table}

\subsection{Ionisation-parameter distribution}
\label{longioni}
\begin{figure*}
\includegraphics[width=18cm]{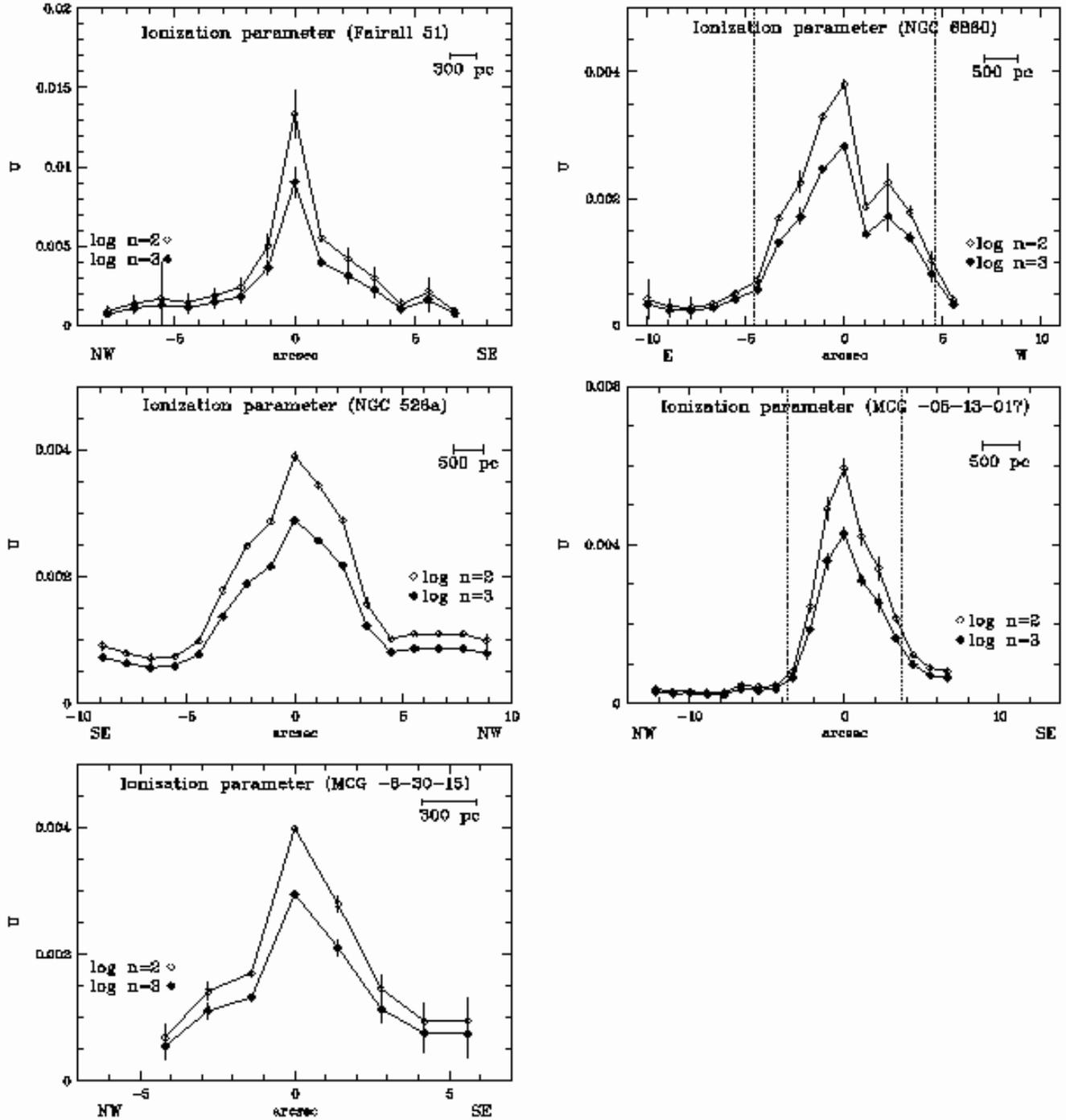}
\caption[]
{\label{ioni1} \small 
Ionisation parameter
derived from [\ion{O}{ii}]/[\ion{O}{iii}] ratio
as a function of the distance from the nucleus 
for 
Fairall\,51, NGC\,6860, NGC\,526a, 
MCG\,-05-13-017, and MCG\,-6-30-15 (open symbols: $n_H$ = 100\,cm$^{-3}$,
filled ones: $n_H$ = 1000\,cm$^{-3}$). 
The edge of the NLR as determined from the
  diagnostic diagrams is indicated by dotted lines (NGC\,6860 and MCG\,-05-13-017).}
\end{figure*}

The line ratio
[\ion{O}{ii}]$\lambda$3727\,\AA/[\ion{O}{iii}]\,$\lambda$5007\,\AA~can be used
to estimate the value of the ionisation parameter $U$ 
[e.g.~\citet{pen90, kom97}]. Here, we followed the method
described in paper I.

The ionisation parameter peaks at the optical nucleus and
decreases with distance. 
In NGC\,6860, a secondary distinct peak is visible.

We fitted a 
power-law function $U_{\log (n_e) = 2, obs}$
= $U_{0}$$ (\frac{R}{R_0})^{\delta}$ to the observed ionisation parameter (Table~\ref{fitioni})
(with $R_0$ = 100 pc
from the nucleus; Table~\ref{fitioni}). 
We include only data points within the NLR.
 $\delta$ ranges
between -0.6 and -1. 

As for the electron density, Seyfert-1 galaxies tend to show a steeper slope than Seyfert-2 galaxies
($\delta_{\rm 5 Sy1} \sim -0.81\pm0.07$ versus $\delta_{\rm 2 Sy2} \sim -0.51
\pm 0.08$; paper II).
However, first the individual scatter is rather large and second,
only two Seyfert-2 galaxies were included in this comparison.

\begin{table} 
\begin{minipage}{80mm}
\caption[]{\label{fitioni} Fitting parameters of ionisation-parameter distribution\footnote{A linear least-squares fit was applied with $\log$$U_{\log (n_e) = 2, obs}$ = $\delta \cdot
\log R/R_0 + \log$$U_{0}$. $U_{0}$ corresponds to the value at $R_0$ = 100 pc distance
from the centre. The number of data points included in the fit is
given in column 2 (= half the number of averaged values from both sides of
the nucleus). For those objects which show a transition between line
ratios typical for AGNs and \ion{H}{ii}-region like ones in the diagnostic
diagrams, determining the size of the NLR, only data points within the NLR
were included (NGC\,6860 and MCG\,-05-13-017).  For
Mrk\,915, the [\ion{O}{ii}] line was not covered by the observations.
}}
\begin{center}
\begin{tabular}{lccc}
\\[-2.3ex]
\hline
\hline\\[-2.3ex]
\multicolumn{1}{c}{Galaxy} & Data Points & $\delta$ & $\log U_{0}$\\[0.25ex]
\hline\\[-2.3ex]
Fairall\,51 &  6     & -0.81$\pm$0.12  & -1.9 \\
NGC\,6860 & 4        & -0.62$\pm$0.25 & -2.2 \\
NGC\,526a & 8        & -0.69$\pm$0.10  & -2.1 \\
MCG\,-05-13-017 & 3 & -1.01$\pm$0.26 & -1.9 \\
MCG\,-6-30-15\footnote{Correction with reddening determined from continuum slope} & 3 & -0.90$\pm$0.20 & -2.7\\[0.1ex]
\hline\\[-2.3ex]
\end{tabular}
\end{center}
\end{minipage}
\end{table}

\subsection{Velocities}
\label{longvel}
We derived the NLR line-of-sight velocity curve by taking the average of
velocity centroids derived by fitting Gaussians to H$\alpha$ and
[\ion{N}{ii}]
as well as the [\ion{O}{iii}] emission lines. In addition, given the high S/N ratio of our
spectra, we were able to trace the stellar rotation curves from
Gaussian fits to
the stellar absorption line \ion{Ca}{ii} K for two objects
(before subtraction of the stellar template) throughout the whole region as
these lines are not blended with emission lines.
The results (with spectral lines used for individual objects indicated)
are shown in Fig. 11. We estimated the uncertainty in determining
the velocity peaks to $\sim 20$ km/s for both the emission and absorption
lines. Note that for Fairall\,51, the [\ion{N}{ii}] emission line
is blended by the strong H$\alpha$ emission and we show the velocity curve
derived from the H$\alpha$ peak alone. 
For MCG\,-6-30-15, the H$\alpha$ and [\ion{N}{ii}] line are strongly blended and
no separation is possible.

As pointed out in paper II, the
interpretation of the NLR velocity curves can be quite complex
and requires modelling of the 3D structure which is beyond the scope
of this paper.  
Here, we limit ourselves to point out that
all the galaxies show large-scale velocity gradients across their NLR.
Based on our preliminary modelling, we believe that to the zeroth order,
they can be explained by rotation in at least 4 galaxies:
Fairall\,51, NGC\,6860, 
Mrk\,915, and MCG\,-05-13-017. The situation is more
complex in NGC\,526a and 
MCG\,-6-30-15. We will present detailed modelling
of velocity fields in a separate paper.

\begin{figure*}
\includegraphics[width=18cm]{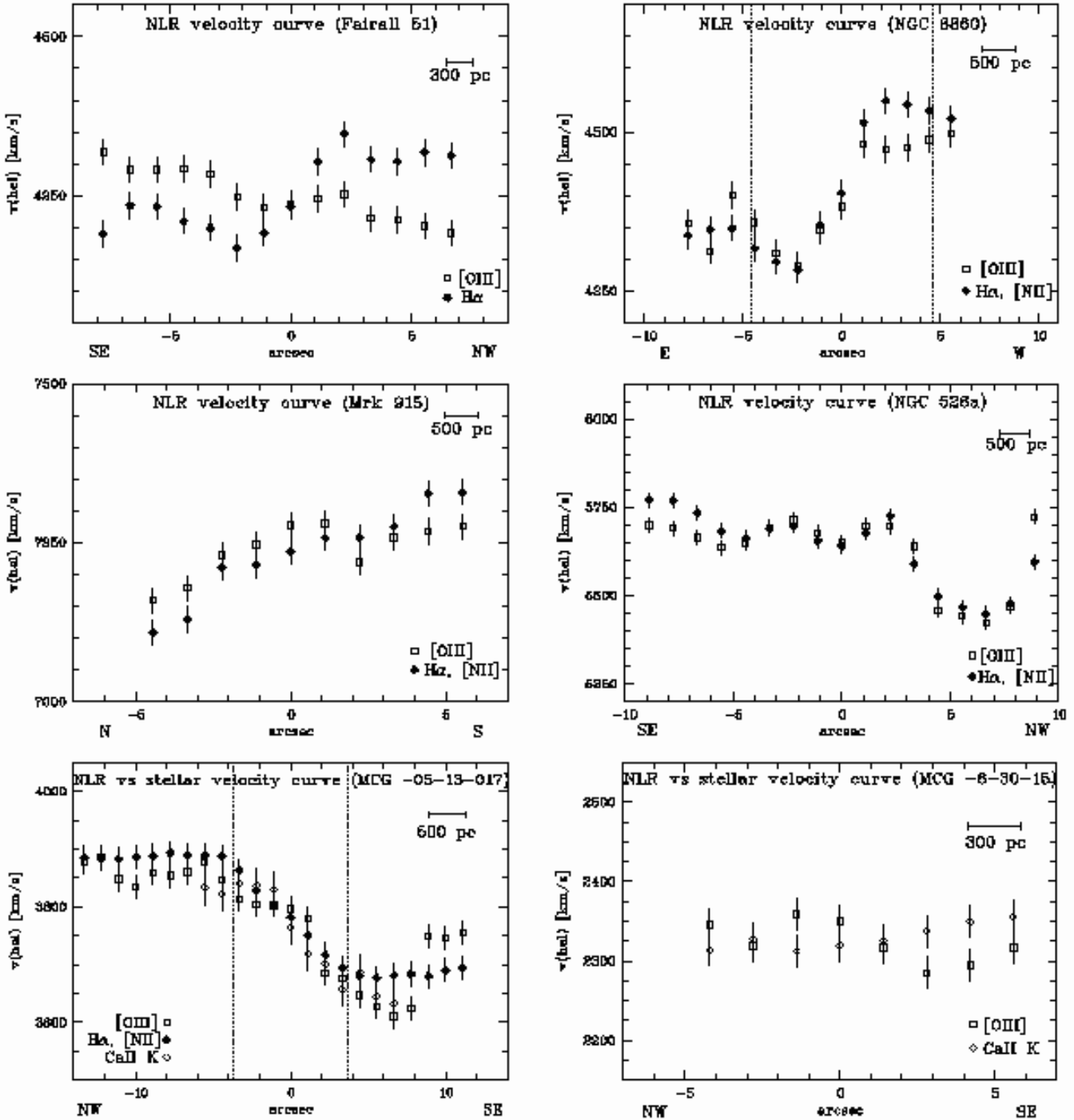}
\caption[]
{\label{vel1} \small
Velocity fields of 
Fairall\,51, NGC\,6860, Mrk\,915, NGC\,526a, MCG\,-05-13-017,
and MCG\,-6-30-15. 
The velocities of
the NLR were derived from the average value of the peak wavelengths
of the H$\alpha$ and [\ion{N}{ii}] emission lines (filled diamonds), with the
exceptions of  Fairall\,51 and MCG\,-6-30-15 where H$\alpha$ and [\ion{O}{iii}] 
were used, respectively.
The [\ion{O}{iii}] velocities are also shown for all objects (open squares).
The stellar 
velocities were determined from the \ion{Ca}{ii}\,K absorption
line ``peak wavelength'' as seen in the ``raw'' spectrum (open diamonds) if
visible at a good S/N.
The edge of the NLR as determined from the
  diagnostic diagrams is indicated by dotted lines (NGC\,6860 and MCG\,-05-13-017).}
\end{figure*}

\section{Conclusions}
We use high-sensitivity spatially-resolved spectra, obtained along the extended
[\ion{O}{iii}] emission with the VLT and the NTT, to study the BLR and NLR
of six Seyfert-1 galaxies.

The nuclear spectra reveal the typical strong NLR emission from oxygen at
different ionisation states, lines from ionised nitrogen and sulphur,
as well as Balmer lines. In addition, broad H$\alpha$ emission is seen
in all objects, broad H$\beta$ emission in all but NGC\,526a,
classifying the latter as Sy1.9.
In most objects, high-excitation iron lines are
seen in the central spectra, originating from the
powerful and hard ionisation source in the centre. High-ionisation
emission lines as well as those with high critical densities tend to be
stronger in Seyfert-1 galaxies. We determine the electron temperature and ionisation parameter in the optical nucleus and find that they
are in general higher in type-1 Seyferts than in type 2s. 

From the continuum luminosity at 5100\AA~as well as the FWHM of
the broad H$\beta$ line, we estimate BH masses and compare them to those
derived from $\sigma_\star$ (as taken from literature).
The Seyfert-1 galaxies in our sample cover a BH mass range of 
$\sim$1 $\cdot$ 10$^{7}$ to $\sim$1 $\cdot$ 10$^{8}$ M$_{\odot}$.

In
addition to the Seyfert-2 galaxies NGC\,1386 and NGC\,5643
already discussed in paper I \& II,
we observe a transition of emission-line ratios from the central AGN region
to \ion{H}{ii} region in two objects (NGC\,6860 and MCG\,-05-13-017),
when plotting line-ratios from our spatially resolved spectra
in diagnostic diagrams.
This transition occurs at a distance of several arcseconds on both sides
of the optical nucleus and is observed in
all three diagnostic diagrams, i.e.~including the second diagnostic diagram
involving the [\ion{O}{i}] emission line.
The most probable explanation for 
this transition is that the stellar ionisation field
starts to dominate that of the AGN. This conclusion is supported
by \texttt{CLOUDY}  photoionisation modelling presented in paper I.
We are thus able to determine the radius of the NLR in these objects
to 700-1500\,pc
independent of sensitivity and excluding
[\ion{O}{iii}] contamination from circumnuclear starbursts.
In former spectroscopic studies, the observed [\ion{O}{iii}] has often been
attributed to the extended NLR. We can show that at least part of this
``extended NLR'' emission is actually predominantly powered by \ion{H}{ii} regions and
that only the central few arcseconds are indeed gas photoionised by the AGN.
For the other four objects, all line ratios
fall in the AGN regime in all three diagnostic diagrams. 
Thus, the determined NLR size (700-3300\,pc)
is a lower limit, limited by either the S/N of our data or the lack of a
strong surrounding stellar ionisation field.

We derive reddening,
surface brightness, electron density, and ionisation parameter within
the NLR as
a function of projected distance from the nucleus.
Both electron density and ionisation parameter decrease with radius.
In general, the decrease is faster in Seyfert-1 galaxies than in type 2s.

We discuss the results for each object individually and compare
them to literature data (Appendix).

Comparing the results presented here to those of six Seyfert-2 galaxies
from paper II shows that both types have in general
similar NLR properties. However, there are differences in emission-line
strength as well as in the
observed slope of spatially varying parameters. The origin
of these differences will
be discussed in a subsequent paper 
on the basis of the unified model taking into account
line-of-sight integrations.

Applying the methods presented here to a larger sample of Seyfert galaxies
will help to measure the NLR size and thus
verify the NLR size-luminosity relation \citep{ben02}.
Our results have shown that although  [\ion{O}{iii}] imaging is less time
intensive than the spectroscopic method, it often yields an either
too small NLR size in case of low sensitivity or a
too large NLR size in case of circumnuclear \ion{H}{ii} regions contributing
to the [\ion{O}{iii}] emission.

\begin{acknowledgements}
We thank the anonymous referee for valuable suggestions.
N.B. is grateful for financial support by the ``Studienstiftung
des deutschen Volkes''. B.J. acknowledges the support of the Research 
Training Network ``Euro3D-Promoting 3D Spectroscopy in Europe''
(European Commission, Human Potential Network Contract No. 
HPRN-CT-2002-00305) and of the Institutional Research Plan No. AV0Z10030501 of the
Academy of Sciences of Czech Republic. M.H. is supported by ``Nordrhein-Westf\"alische
Akademie der Wissenschaften''.
We thank Pierre Ferruit for providing and helping
us with the \texttt{fit/spec} line-fitting tool.
Henrique Schmitt was so kind to provide the continuum-subtracted
HST [\ion{O}{iii}] images of several Seyfert galaxies in this sample.
This research has made use of the NASA/IPAC Extragalactic Database (NED), 
operated by the Jet Propulsion Laboratory, Caltech, under contract with the NASA.
\end{acknowledgements}

\Online

\appendix
\section{Comments on Individual Objects}
\label{comments}
We searched the available literature for all objects
in our sample and
here summarise the most important results in comparison with our study.
Note that the velocity fields will be discussed in detail
when comparing them to those derived from modelling in a subsequent paper.

\subsection{Fairall\,51}
\label{f51}
Fairall\,51 
is known for its high polarisation, noticed already
by \citet{mar83}, \citet{tho88}, and \citet{bri90}.
Fairall\,51 is a barred spiral with 
the bar extending to about 12\arcsec~on both sides of the nucleus in
north-south direction \citep{wes78}. 
\citet{schmid01} investigated the origin of the polarisation and the geometry of
the system using VLT spectropolarimetry.
While the AGN continuum and the broad lines reveal a practically identical
amount of intrinsic polarisation (5\% in the red up to 13\% in the UV), 
the narrow lines are unpolarised or show only little intrinsic polarisation.
The spectrum is much redder in total flux than in polarised flux.
These Seyfert 2-like polarisation characteristics indicate that
the nucleus is partially obscured with the scattering region located
far from the BLR and the continuum source. 
This specific geometry allows to study the BLR both directly and via scattering.
From the indistinguishable line profiles in polarised and total light,
\citet{schmid01} conclude that the velocity field of the BLR is spherically
symmetric.

Fairall\,51 is the object with the strongest iron emission lines 
relative to H$\beta$ seen in the nuclear spectrum
in our sample: Compared to the average values determined
from the four other Seyfert-1 galaxies from Table~\ref{lineratio1}, 
the [\ion{Fe}{vii}]\,$\lambda$5721\,\AA/H$\beta$ flux ratio is more
than twice as high in Fairall\,51, the
[\ion{Fe}{vii}]\,$\lambda$6087\,\AA/H$\beta$ flux ratio is roughly three times as
high, and the [\ion{Fe}{x}]\,$\lambda$6375\,\AA/H$\beta$ flux ratio is $\sim$
four times higher than the average value.
We detect [\ion{O}{iii}] emission (S/N $>$ 3) out to a distance of 9\arcsec~from
the nucleus (Table~\ref{tablediag}), i.e.~three times larger than the maximum
extension of the HST [\ion{O}{iii}] image from \citet{sch03a} ($\sim$2\farcs65). 
Emission-line ratios out to $\sim$8\arcsec~indicate ionisation by
the central AGN source, although the outer 6-7\arcsec~north-west of the
nucleus (marked as E/F in the
diagnostic diagram) show line ratios at the border between those typical for
AGNs and those expected for LINERs (Fig.~\ref{diag3}). 
We classify the inner
14\arcsec~as NLR (Table~\ref{tablediag}). 

The reddening of the central BLR, 
deduced from the broad H$\alpha$/H$\beta$ ratio
of $\sim$4.4, is comparable to the central NLR reddening 
($E_{(B - V),\rm broad} \sim
0.45$\,mag). 
It indicates that the dust is 
external to both emission line regions.
We interpret
deviations of the broad Balmer decrement from the recombination value as
reddening indicator, but note that
optical depth effects could lead to similar results [e.g. \citet{net75}].
However, the fact that both, NLR and BLR show the same
amount of reddening in Fairall\,51 (and several of the other galaxies
of our sample) strongly hints at a common absorber affecting
both regions in the same way, rather than optical depth
effects in the BLR.

In the central spectra, the broad emission lines in both H$\alpha$ and H$\beta$
clearly dominate the profile, confirming the classification of Fairall\,51 as
pure Seyfert-1 galaxy.

In the same region where we find line ratios approaching
that expected for LINERs, we observe an enhanced surface brightness in the
[\ion{O}{iii}] and H$\alpha$ emission as well as the
continuum (Fig.~\ref{lum1}). Moreover, two density peaks on both sides of the nucleus
roughly coincide with the enhanced surface brightness and the north-western region in which
line ratios close to LINER ones are observed 
(Fig.~\ref{density1}). Also the
ionisation parameter in Fairall\,51
reveals a small secondary peak at $\sim$6\arcsec~north-west of the nucleus 
(Fig.~\ref{ioni1}). Is the match of the increased electron density, ionisation parameter, surface
brightness as well as the line ratios close to LINER type at 5-6\arcsec~to the
north-west of the nucleus pure coincidence
or does it hint e.g.~shocks occurring at these distances? Unfortunately, there
is no published information on the radio source in Fairall\,51 allowing to
discuss a possible jet-NLR interaction. Alternatively,
the bar (p.a.~$\sim 180$\degr) may influence the NLR, resulting in the
observed properties on both sides of the centre (p.a.$_{\rm obs}
\sim$ 160\degr).

\subsection{NGC\,6860}
\label{ngc6860}
The luminous infrared galaxy
NGC\,6860 shows a composite nature of a Seyfert-1 nucleus embedded in a dusty
star formation environment \citep{lip93}. While NED classify NGC\,6860 as Seyfert 1,
\citet{cid98} and \citet{lip93} find emission lines typical for a Seyfert 1.5.
As the narrow components in both H$\alpha$ and H$\beta$ which are superimposed on the
broad components are clearly visible in our central spectra, we agree with
\citet{lip93} to classify NGC\,6860 as Sy1.5.

\citet{lip93} studied NGC\,6860 
in detail, presenting optical imaging as well as
optical and near-infrared spectroscopy.
The H$\alpha$+[\ion{N}{ii}] image shows 
a bright emission-line region associated with the AGN activity
and a circumnuclear ring of star formation with clear signs
of young stars in the spectrum. 
The [\ion{O}{iii}] image reveals emission extended by 10\arcsec~in east-west
direction, perpendicular to the
direction of the bar (p.a.$_{\rm bar}$~$\sim$ 13\degr). 
\citet{lip93} use templates from \citet{bic88} to fit the stellar population,
finding that an S3 template represents the nuclear stellar population, while
the rather young population of an S6 template fits the inner star formation ring.
The bar reveals a typical spectrum of an old stellar population.
A high electron temperature ($T_{\rm e, obs}$ $\sim 35000$ K) is observed in the nuclear region.
NGC\,6860 is classified as a typical 
intermediate case between AGN activity
completely dominating the energy input
and starburst galaxy where the ionising continuum is spread over a much larger
star-forming region \citep{lip93}. This is strengthened by line ratios in
diagnostic diagrams which locate the galaxy in a transitional zone between
areas occupied by AGNs and \ion{H}{ii} regions.

Besides MCG\,-05-13-017, NGC\,6860 is the other galaxy in our
sample which shows, in all three diagnostic diagrams,
a clear transition between central line ratios falling in
the AGN regime and outer ones typical for \ion{H}{ii} regions 
(Fig.~\ref{diag1}).
While we detect [\ion{O}{iii}] emission at a S/N $>$ 3 
out to a distance of $r \sim$ 10\arcsec~in east-west direction, comparable to the
maximum extension of the [\ion{O}{iii}] groundbased image of \citet{lip93}, 
the emission beyond 5\arcsec~can be attributed to circumnuclear star forming
regions (Table~\ref{tablediag}). 
Our results confirm the classification of NGC\,6860 as an intermediate case
between AGN activity and starburst galaxy \citep{lip93}. While \citet{lip93}
find line ratios in the transitional zone between Seyfert galaxies and
\ion{H}{ii} regions, we are able to trace the radial varying ratios and
determine the radius where the transition takes place in 
all three diagnostic diagrams. Showing that the transition occurs at the same
distance in the second diagnostic diagram is
essential to exclude spatial variations of physical parameters 
resulting in a transition of line ratios from the AGN to the
\ion{H}{ii}-region 
regime despite an intrinsic AGN photoionising source (paper I).
Thus, we can determine the NLR size of NGC\,6860 to a radius of $r \sim
5$\arcsec~(observed along a p.a.~of 85\degr), i.e.~the AGN radiation field
dominates over the stellar one in the inner 10\arcsec.
The maximum [\ion{O}{iii}] extent we observe is four times larger than what is
observed in the HST [\ion{O}{iii}] image by \citet{sch03a} 
($d \sim 20$\arcsec~versus $d\sim 5$\arcsec).

In the centre, we find a high electron temperature of $T_{\rm e, obs}$ $\sim 36325\pm250$\,K,
in agreement with the results of \citet{lip93} (Table~\ref{result}).
The reddening values vary rather randomly. Moreover, there is no
significant difference between the reddening in the NLR and that in the
surrounding \ion{H}{ii} regions (Fig.~\ref{reddening1}). 
The BLR shows a reddening comparable to that of the NLR
(in the central 2\arcsec: 
H$\alpha_{\rm broad}$/H$\beta_{\rm broad}$ $\sim$ 5.5, $E_{B - V} \sim 0.6$\,mag).
As in Fairall\,51, it indicates dust in the host
galaxy.

\subsection{Mrk\,915}
\label{mrk915}
Mrk\,915 is a highly-inclined ($i \sim 80$\degr) Seyfert-1 galaxy with a
companion at 126\arcsec~to the south-east \citep{kee96}.

\citet{goo95} observes flux changes in the broad H$\alpha$ line which has
almost disappeared in 1993 compared to spectra obtained nine years earlier
(their Fig. 3). If
this change can be explained by an increase in reddening, then $\Delta
E_{(B-V)} \ge 0.53$\,mag. 
The broad H$\alpha$ emission we find in our nuclear spectra observed in
September 2004 
is comparable to that seen by \citet{goo95} 
in 1984, i.e.~it is significantly higher than that
observed in 1993 ($F_{\rm H\alpha, 1984} = 637 \cdot
10^{-15}$\,erg\,s$^{-1}$\,cm$^{-2}$, $F_{\rm H\alpha, 1993} \le 223 \cdot
10^{-15}$\,erg\,s$^{-1}$\,cm$^{-2}$, $F_{\rm H\alpha, 2004} = 649 \cdot
10^{-15}$\,erg\,s$^{-1}$\,cm$^{-2}$). 
If this change can be attributed to dusty clouds,
they have high transverse velocities and must thus be close to, but outside of, the bulk
of the BLR itself \citep{goo95}.
As both the broad and the superimposed narrow components in H$\alpha$
and H$\beta$ can be easily recognised in the nuclear spectra, we classify
Mrk\,915 as Seyfert-1.5 galaxy rather than Sy1. 
Calculating the reddening of the
  BLR in the central 2\arcsec~yields a reddening value comparable to that of the
  NLR, indicating that, at the moment,  there is no significant dust amount between the 
  broad and narrow Balmer lines
 (H$\alpha_{\rm broad}$/H$\beta_{\rm broad}$ $\sim$ 5; $E_{B - V,\rm broad} \sim$ 0.5\,mag).

A groundbased H$\alpha$+[\ion{N}{ii}] image were taken by \citet{col96},
showing extended emission along the major axis (p.a.~$\sim$ 168\degr) 
out to a radius of 15\arcsec~in south-eastern direction.
Along a p.a.~of 5\degr, the H$\alpha$+[\ion{N}{ii}] emission extends out to
$\sim$10\arcsec.

The HST [\ion{O}{iii}] image of \citet{sch03a} reveals irregular emission with
a major extent of 4\farcs1 along p.a.~= 5\degr~and 2\farcs6 in the
perpendicular direction. 
We detect [\ion{O}{iii}] emission at a S/N $\ge$ 3
out to a radius of $r \sim$12\arcsec, i.e.~three times further out than the HST
[\ion{O}{iii}] image (Table~\ref{tablediag}). 
The [\ion{O}{iii}] extension obtained from our spectra
is comparable to the $r \sim 10$\arcsec~extension seen in the groundbased
H$\alpha$+[\ion{N}{ii}] image at a p.a.~of 
5\degr~[\citet{col96}, see their Fig. 2h].
Emission-line ratios at S/N $>$ 3 were traced out to a distance of $r \sim$ 6\arcsec~from the
nucleus. Within this region, all ratios are typical for AGN ionisation and we
thus classify the emitting region as NLR 
(Table~\ref{tablediag}; Fig.~\ref{diag3}).

\subsection{NGC\,526a}
\label{ngc526a}
NGC\,526a is the brighter member of a strongly interacting pair of galaxies.
The Seyfert type is discussed controversially: \citet{mul96a} and \citet{whi92}
list NGC\,526a as Sy2, while RC3 classify it as Sy1.5. 
\citet{win92} discuss the presence of broad H$\alpha$ and the absence of
broad H$\beta$, suggesting a classification of Sy1.9.
Our spectra do not show signs of broad H$\beta$ emission but a broad H$\alpha$
component is clearly visible in the central 2\arcsec. 
Thus, we agree with
\citet{win92} in classifying NGC\,526a as Sy1.9.
The central (narrow) H$\alpha$/H$\beta$
ratio of $\sim$ 4.1 we observe is slightly higher
than the value reported by \citet{win92} ($\sim3$) but significantly lower
than the reddening of the BLR: We tried to fit any spurious underlying broad H$\beta$
flux and find, in agreement with \citet{win92}, a high reddening of the BLR
with H$\alpha$/H$\beta$ $\ge$ 10, corresponding to $E_{B - V, {\rm broad}}$
$\ge$ 1.26\,mag, larger than that of the NLR.
The BLR reddening implies an absorbing column density of $N_{\rm H}$
$\sim$ 7.4 10$^{21}$\,cm$^{-2}$  for a Galactic gas/dust ratio
(and in the absence of optical depth effects),
which is similar to the amount of X-ray absorption [$N_{\rm H} \sim 10^{22}$
\,cm$^{-2}$; \citet{lan01}].

Groundbased emission-line images in [\ion{O}{iii}] and
H$\alpha$+[\ion{N}{ii}] are presented by \citet{mul96a}.
The images reveal extended line emission out to $r \sim$ 10\arcsec~in north-west/south-east
direction, in approximately the direction of the companion galaxy.
Excitation maps show evidence of a bi-conical shape of the highest excitation gas.

We detect extended [\ion{O}{iii}] emission at a S/N $>$ 3 out to a radius of 
$\sim$20\arcsec~($\sim$7240\,pc) from the nucleus (Table~\ref{tablediag}), 
i.e.~twice as far as seen
in the groundbased emission-line image of \citet{mul96a}.
The line ratios needed for
diagnostics and the derivation of physical parameters where determined out to
9\arcsec~($\sim$3260\,pc) to the north-west and south-east from the nucleus. All
emission-line ratios within this region indicate AGN ionisation
(Table~\ref{tablediag}; Fig.~\ref{diag3}). 

\subsection{MCG\,-05-13-017}
\label{mcg5}
According to \citet{mul96a}, the Seyfert-1.5 galaxy
MCG\,-05-13-017 is a strongly perturbed galaxy,
showing extended [\ion{O}{iii}] emission out to a radius of $\sim$11\arcsec~to 
the south-east around an emission component concentrated in the nucleus. 
The [\ion{O}{iii}] emission extends roughly along the
host galaxy major axis (p.a.~= 160\degr~with $i = 54$\degr; RC3).
The groundbased H$\alpha$ image
reveals more symmetrically distributed emission, with the strongest
off-nuclear emission within the [\ion{O}{iii}] extension. From the excitation
map, \citet{mul96a} concludes that not all of the gas within the
south-eastern extension is of high excitation.

The most extensive spectroscopic study of the extended NLR of MCG\,-05-13-017 has been
performed by \citet{fra00}.
Emission-line fluxes were measured along p.a.~= 158\degr~out to
17\arcsec~from the nucleus. The [\ion{O}{ii}]/[\ion{O}{iii}] ratio increases with distance from the
nucleus, indicating a decreasing ionisation parameter.
Comparing these results with groundbased images of
\citet{mul96a}, \citet{fra00} suggest that the nuclear continuum ionises the
gas in the disk along p.a.~= 158\degr, giving rise to the cone-shaped region
observed in [\ion{O}{iii}].

We detect [\ion{O}{iii}] emission with a S/N $>$ 3 out to a radius of
17\arcsec~from the photometric centre (Table~\ref{tablediag}). This is
significantly larger than the $d \sim 2$\arcsec~total extension seen in the HST
snapshot image and also larger than what has been determined from 
the groundbased image of  \citet{mul96a} ($r \sim 11$\arcsec).
The excitation map of \citet{mul96a} already indicates that not all of the gas
is of high excitation and the diagnostic diagrams in our study can confirm
this finding (Fig.~\ref{diag1}). 
Moreover, we can show that only the central $\pm$3\arcsec~show line
ratios typical for AGN ionised gas (marked with 0, A/a, B/b, C,c, respectively
in Fig.~\ref{diag1})
and that the spectra out to $r \sim$
11\arcsec~north-west and out to $r \sim$ 6\arcsec~south-east
have line ratios falling in the regime covered by \ion{H}{ii} regions
(D/d to F/k).
The outer line ratios coincide with the position of \ion{H}{ii} regions
defined by \citet{tsv95}: They classify 38 \ion{H}{ii} regions 
in their groundbased 
H$\alpha$+[\ion{N}{ii}] image of MCG\,-05-13-017, distributed in a cloud around the
nucleus with distances between 3-18\arcsec~from the nucleus.
Along the p.a.~of 140\degr~as used in our long-slit
observations, \ion{H}{ii} regions start at a distance of
4\arcsec~from the nucleus and extend out to 11\arcsec.
While \citet{fra00} attribute emission out to $\sim$17\arcsec~from the nucleus
to the extended NLR, we can show that only the central $\pm$3\arcsec~are gas
ionised by the central AGN.

MCG\,-05-13-017 is the object with the highest central electron temperature in our
sample ($T_{\rm e, obs}$ $\sim$ 52500$\pm$3000\,K: Table~\ref{result}). 
The ionisation parameter is also high
in the centre.
These values reflect in high flux ratios of high-excitation lines
(Table~\ref{lineratio1}):
The nuclear spectrum shows strong
[\ion{Ne}{iii}]\,$\lambda$3869\,\AA~emission, even 
exceeding that of [\ion{O}{ii}]\,$\lambda$3727\,\AA; it is 
the highest flux ratio [\ion{Ne}{iii}]/H$\beta$ ($F_{\rm dered}$  = 3.43) 
observed in our sample. Also the ratios
[\ion{O}{iii}]\,$\lambda$5007\,\AA/H$\beta$ ($F_{\rm dered}$  = 16.3; $R_{\rm dered}$ =
[\ion{O}{iii}]\,$\lambda$4959\,\AA+5007\,\AA/H$\beta$ $\sim$ 21.7) and 
[\ion{O}{i}]\,$\lambda$6300\,\AA/H$\beta$ ($F_{\rm dered}$  = 0.79) are
the highest ones of our sample. Additionally, strong coronal lines are
observed.

The broad Balmer decrement H$\alpha_{\rm
  broad}$/H$\beta_{\rm broad}$ of 5.7 indicates a slightly higher reddening of the
BLR with respect to the central NLR ($E_{B - V, \rm broad} \sim 0.7$\,mag), possibly due to dust
in between the two regions.
The electron density is highest in the centre ($n_{\rm e, obs}$ $\sim$ 2460$\pm$55\,cm$^{-3}$; Table~\ref{result}). 
This value is roughly 2.5 times higher than the value
reported by \citet{rod00} which is approximately the factor that we took into
account when correcting
with the observed central temperature of $T_{\rm e, obs}$ $\sim 52500$\,K\footnote{$n_e (\rm {T}) 
= n_e ({\rm [SII]\,ratio}) \cdot \sqrt{(T/10000)}$}. 

The ionisation parameter is  peaked in the nucleus and rapidly
decreases within the NLR  (Fig.~\ref{ioni1}). 
Our observations confirm the results of \citet{fra00} who already
suggest that the ionisation parameter is decreasing, based on the
increased [\ion{O}{ii}]/[\ion{O}{iii}] ratio (observed along a slightly
different p.a.~of $\sim$ 20\degr).

\subsection{MCG\,-6-30-15}
\label{mcg6}
The E/S0 galaxy MCG\,-6-30-15 hosts an X-ray bright 
AGN of Seyfert 1.2 type (NED).
It has been the subject to intense spectroscopic studies in the
X-ray band [e.g.~\citet{iwa96}] and is the best candidate for
harbouring a relativistically broadened iron line [e.g.~\citet{tan95}].

A multiwavelength study has been performed
by \citet{rey97}: In the optical, an extracted host galaxy spectrum 
with Balmer and \ion{Ca}{ii} absorption features
is subtracted to gain the absorption-line
free AGN spectra. The nuclear spectrum reveals 
a strong non-stellar continuum, broad Balmer lines, and narrow permitted and
forbidden lines. High-excitation forbidden lines
(e.g.~[\ion{Fe}{x}],
[\ion{Fe}{xi}]\,$\lambda$7892\,\AA, and [\ion{Fe}{xiv}]\,$\lambda$5303\,\AA)
are clearly displayed.
A significant amount of dust extinction is deduced from optical line
and continuum emission, lying in the range $E_{(B-V)}$ =0.6--1\,mag.
Due to the absence of cold (neutral) absorption in the X-ray spectra,
\citet{rey97} postulate that the dust resides in the so-called warm absorber, 
i.e.~ionised gas that absorbs X-rays produced in the accretion
disk of the AGN. 

VLA observations at both 3.6 and 20\,cm show that
MCG\,-6-30-15 is an unresolved radio source \citep{nag99b}.
HST continuum images and colour maps from \citet{fer00} show a dust lane south
of the nucleus, running roughly parallel to the photometric  major axis of the
galaxy (p.a.~$\sim$ 115\degr; $i \sim$ 60\degr). The south-west side of the
galaxy is found to be systematically redder than the north-east side, with the
central regions redder than the outer ones. 

The VLT long-slit spectra of MCG\,-6-30-15 allow us to trace
[\ion{O}{iii}] emission out to a distance of $\pm$12\arcsec~from the optical
nucleus, i.e.~six times larger than the [\ion{O}{iii}] extension seen
in the HST image of \citet{sch03a}. 
We confirm the existence of high-excitation forbidden lines in the central
spectra as reported by \citet{rey97}. 

MCG\,-6-30-15 is the only galaxy in our sample in which we were not
able to disentangle the broad and narrow Balmer emission lines. 
This is first due to
the classification as Sy1.2 whereas the other Seyfert-1 galaxies (apart
from Fairall\,51) can be classified as either Sy1.5 or Sy1.9, i.e.~a clear
separation of the narrow emission line superimposed on the broad one is
visible. Second, it is also due to the lower resolution of our VLT/FORS1 spectra while all
other type 1s were observed with the higher spectral resolution of the NTT/EMMI.
Thus, we cannot determine the central emission line fluxes relative to
H$\beta$. Moreover, it was difficult to disentangle the
[\ion{O}{iii}]\,$\lambda$4363\,\AA~emission line from the broad H$\gamma$ and
we therefore do not measure the temperature. 
MCG\,-6-30-15 is also the only Seyfert-1 galaxy for which we applied a stellar
template subtraction. This was necessary as strong Balmer and
\ion{Ca}{ii} H\&K absorption lines were seen in the NLR spectra
(Fig.~\ref{figtemplate}), in agreement with the results of 
\citet{bon00} who find an old bulge population dominating the stellar contribution,
with indications of previous bursts of star formation.

The continuum gets steadily redder towards the nucleus
where it reaches the highest reddening value of $E_{B - V} \sim$ 0.3$\pm$0.02\,mag.
This amount of dust extinction is significantly lower than 
what has been determined by \citet{rey97}. They were also not able
to disentangle the broad and narrow emission lines in their low resolution
spectra and use instead the total Balmer decrement.
They derive a range of $E_{B - V}$ $\sim$ 0.6--1\,mag. 
The difference may arise from either additional reddening of the BLR only [as
\citet{rey97} use the total Balmer decrement] or is due to the fact that our reddening 
value is a value relative to the stellar template and simply reflects the
differences of dust extinction between the NLR and the template.
If the extinction is at least partially due to dust in the ISM of the host galaxy
as proposed by \citet{bal03}, both the stellar template and the central
continuum suffer the same extinction which will then not reflect in the reddening
determined by the fitting of the stellar template to the continuum slope of
the NLR spectra.

As the broad Balmer lines are limited to the central
2-3\arcsec,
we are able to measure the narrow
line fluxes in the outer regions to a distance of $\pm$4\arcsec.
We find that the line ratios are typical for AGNs when inserting
them in the diagnostic diagrams.~Given the broad emission lines, 
it is reasonable to believe that the central values also lie in
the AGN regime and that the NLR extends out to at least 4\arcsec~distance from
the nucleus.

The BH mass we estimate (0.8-3 $\cdot$ 10$^7$ M$_{\odot}$) is roughly in
agreement with 
the various measurements of the BH mass presented by \citet{mch05}
(3-6 $\cdot$ 10$^6$ M$_{\odot}$), taking into account the errors.

We used the reddening determined from the continuum slope variation relative
to the stellar template to correct the observed [\ion{O}{ii}]/[\ion{O}{iii}]
ratio throughout the whole NLR.
The thereby derived ionisation parameter may have a greater uncertainty than
using the reddening within the NLR as measured from the Balmer decrement: In
those objects in which we measured the reddening from both the continuum slope and
the narrow H$\alpha$/H$\beta$ value, we observe differences in the reddening
distributions and, moreover, a
smaller reddening of the continuum is observed (paper I \& II). 
Thus, for MCG\,-6-30-15, the absolute 
ionisation parameter values may be overestimated due to a correction by too
low reddening values, but we consider the  deduced general behaviour as reliable:
In paper I, 
we have shown exemplarily for NGC\,1386 that the general distribution of the
ionisation parameter does not change significantly using either
reddening values for correction.

\end{document}